# Square Moiré Superlattices in Twisted Two-Dimensional Halide Perovskites

Shuchen Zhang[1,10], Linrui Jin[2,10], Yuan Lu[3,10], Linghai Zhang[4], Jiaqi Yang[5], Qiuchen Zhao[2], Dewei Sun[2], Joshua J. P. Thompson[6], Biao Yuan[3], Ke Ma[1], Akriti[1], Jee Yung Park[1], Yoon Ho Lee[1], Zitang Wei[1], Blake P. Finkenauer[1], Daria D. Blach[2], Sarath Kumar[2], Hailin Peng[7], Arun Mannodi-Kanakkithodi[5], Yi Yu[3], Ermin Malic[6], Gang Lu[8], Letian Dou[1,2,9]*, Libai Huang[2]*

[1]Davidson School of Chemical Engineering, Purdue University, West Lafayette, Indiana 47907, United States
[2]Department of Chemistry, Purdue University, West Lafayette, Indiana 47907, United States
[3]School of Physical Science and Technology, ShanghaiTech University, Shanghai 201210, China
[4]School of Flexible Electronics (Future Technologies), Nanjing Tech University, Nanjing 211816, China
[5]School of Materials Engineering, Purdue University, West Lafayette, Indiana 47907, United States
[6]Department of Physics, Philipps-Universität Marburg, 35037 Marburg, Germany
[7]Center for Nanochemistry, Beijing Science and Engineering Center for Nanocarbons, Beijing National Laboratory for Molecular Sciences (BNLMS), College of Chemistry and Molecular Engineering, Peking University, Beijing 100871, China
[8]Department of Physics and Astronomy, California State University Northridge, Northridge, California 91330-8268, United States
[9]Birck Nanotechnology Center, Purdue University, West Lafayette, Indiana 47907, United States

[10]Contributed Equally: S. Zhang, L. Jin, Y. Lu
*Corresponding Authors: L. Huang and L. Dou

**Abstract**

Moiré superlattices of two-dimensional (2D) materials have recently emerged as a new platform for studying strongly correlated and topological quantum phenomena[1-9]. 2D organic-inorganic halide perovskites with programmable structures and strongly bound excitons[10-13] are excellent candidates for creating moiré structures featuring square lattices. Moiré flat bands have been predicted in twisted 2D perovskites[14]; however, their experimental realization is absent owing to a variety of synthetic and fabrication challenges, including difficulties in creating an ultra-thin perovskite sheet free from bulky organic ligands, and in transferring and stacking of these thin sheets that have soft ionic lattices[15-19]. Here, we overcome these obstacles via a new synthetic pathway and demonstrate moiré superlattices based on twisted ultra-thin ligand-free 2D halide perovskites. Square moiré superlattices with different periodic lengths are clearly visualized through high-resolution transmission electron microscopy (HRTEM). Angle-dependent transient photoluminescence microscopy and electrical characterizations indicate the emergence of localized bright excitons and charge carriers trapped by a moiré potential near a twist angle of ~10°. Simultaneously, more than one order of magnitude enhancement of photoluminescence intensity is observed in the 10°-twisted perovskite layers system. These phenomena can be attributed to the formation of narrow moiré flat bands confirmed by first-principles and microscopic many-particle calculations. The combination of hydrogen bonds and long-range ionic interaction across at the interface leads to an unusual large magic angle of ~10°, in sharp contrast to twisted bilayers with van der Waals interactions[1,7,14]. The long-range ionic interaction extends the moiré interference beyond the length-scale of multiple unit-cell-thickness. Our findings provide a new tunable family of 2D ionic semiconducting materials for exploring moiré flat bands at room temperature.

**Main**

Moiré superlattices offer an exciting new paradigm in designing materials properties by utilizing spatially varying interlayer interaction to modify electronic band structures. A plethora of quantum phenomena, including superconductivity and Mott insulators, have been realized in magic-angle graphene[1] and twisted bilayers of transition metal dichalcogenides (TMDCs)[8]. Carriers and excitons can be trapped by periodic moiré potentials, which lead to the formation of flat bands and strongly correlated electronic states. The moiré materials design space can be further expanded by external parameters such as electrical gating and mechanical strains. However, the interlayer interaction has been limited to weak and short-range van der Waals (vdW) forces in layered graphene and TMDCs, and therefore energy modulation introduced by moiré patterns is relatively small (on the orders of 10s meV)[9]. The small energy modulation makes these flat bands susceptible to thermal fluctuations and disorders. As a result, thus far, moiré flat band physics has been almost exclusively observed at low temperatures.

Interlayer interactions beyond vdW are desirable for increasing the depth of energy modulation to realize room-temperature moiré materials[20-21]. However, a major obstacle is fabricating large two-dimensional (2D) non-vdW materials with controlled thickness. An excellent candidate is 2D halide perovskites with ionic bonding to enable stronger interlayer coupling. The perovskite crystal structures can also open up an intriguing square lattice geometry with moiré flat bands[14]. This novel geometry enables the investigation of 2D electronic Hamiltonians, such as Lieb lattices[22], and sets the 2D perovskites apart from other vdW materials. Unfortunately, current approaches[23-28] failed to produce atomically-flat, ultrathin, and ligand-free large-area 2D perovskite crystals with well-controlled twist angles that are necessary for the construction of moiré superlattices. The organic ligands are usually too bulky to allow electronic coupling between the adjacent inorganic layers[29-32] in Ruddlesden–Popper (RP) phase 2D halide perovskites ($L_2A_{n-1}Pb_nX_{3n+1}$, where $L^+$ is a long-chain alkyl or aromatic ammonium cation, $A^+$ is a small cation, $X^-$ is a halide, and n is an integer)[33-36]. An alternative method is to reduce the dimensionality of three-dimensional (3D) $APbX_3$ perovskites via suppressed growth along one specific crystallographic direction (e.g., <001>)[37-41], but it lacks precise control on the thickness and lateral size [36-38]. Furthermore, owing to the ionic nature of the perovskite lattice, neither mechanical transfer nor growth of ultrathin $APbX_3$ on a polymer-based flexible substrate has been realized, both of which are key techniques to stack two thin layers with a controlled twist angle.

In this work, we overcame these challenges to construct twisted perovskite layers (TPLs) by developing a new synthesis method for ligand-free 2D $APbX_3$ perovskites (A = methylammonium [MA] or formamidinium [FA]; X = Br or I) whose thickness can be lowered to 2 nm and whose lateral size can reach over 10 micrometers on arbitrary substrates. The artificially twisted structures feature square moiré patterns and can host moiré flat bands near a twist angle of ~10°. Time-resolved optical microscopy measurements revealed exciton localization by the moiré potential and enhanced exciton emission from the flat bands. These results present exciting opportunities for twisted perovskite structures as a new room-temperature moiré materials platform with the realization of flat bands with long-range ionic interactions.

**Ligand-free 2D $APbX_3$ synthesis and reaction mechanism**

We first demonstrate a new synthesis method to obtain ultra-thin and ligand-free 2D $APbX_3$ on arbitrary substrates, which is crucial for fabricating artificially twisted structures. **Figure 1a** together with Extended Data Fig. 1 shows the schematic of our process to realize the ultrathin 2D $APbX_3$ perovskites. Briefly, directly grown or exfoliated ultrathin 2D RP phase lead halide perovskites (a few nanometers thick) were used as a template and then an 'equilibrium solution', formed by PbX$_2$/AX in isopropanol (IPA) (Supplementary Fig. 1-3), was applied to convert the ultrathin RP phase into the desired APbX$_3$ phase[42-43]. In the equilibrium solution, PbX$_2$ has limited solubility (Table S1) and AX and PbX$_2$ spontaneously reacted into ABX$_3$ at the solid-liquid interface. A small amount of [PbX$_6$]$^{4-}$ forms in the solution, which is crucial for the successful conversion. When the equilibrium solution is dropped on the layered RP phase thin crystals, the existence of [PbX$_6$]$^{4-}$ impedes the dissolution of inorganic layers of [A$_{n-1}$Pb$_n$X$_{3n+1}$]$^{2-}$. This conversion reaction is triggered first at the edge and propagates into the inner part of the crystal (Supplementary Fig. 4). Driven by the concentration gradient, the L$^+$ cations are then released into the IPA solution while A$^+$ cations diffuse in between the layers. Meanwhile, excess I$^-$ anions also enter the solution, facilitating the inorganic layers to merge together through sharing iodine bridge and further leading to the formation of non-layered APbX$_3$ perovskites (**Fig. 1a**). Extended Data Fig. 2 indicates that there is no change in crystal shape before and after the conversion reaction. The reduced thickness observed from atomic force microscope (AFM) images (Supplementary Fig. 5) and the emergence of new peaks at ~14° and ~28° from X-ray diffraction (XRD) (Supplementary Fig. 6) after the reaction both validate the successful conversion to ABX$_3$ structure.

Precise control of the concentration of [PbX$_6$]$^{4-}$ and A$^+$ (i.e., the ratio of PbI$_2$/AX) in the equilibrium solution is crucial for achieving high-quality ABX$_3$ structure using 2D RP perovskites as a template (equation 1). If the PbI$_2$/AX ratio is too high (AX deficient), [PbX$_6$]$^{4-}$ concentration is too low to stabilize the inorganic framework in perovskites, leading to the dissolution of precursors by decomposing into [PbX$_6$]$^{4-}$ (Supplementary Fig. 7-8), as shown in equation 2. On the other hand, when the ratio of PbI$_2$/AX is too low, the halide anions (X$^-$) from excess AX break the bridge between octahedral units to form free [PbX$_6$]$^{4-}$, leading to the disappearance of 2D RP perovskite template (Supplementary Fig. 9-10), described by equation 3. The optimal PbX$_2$/AX molar ratio in the equilibrium solution is found to be around 1:3 (Supplementary Fig. 11 and Table S2).

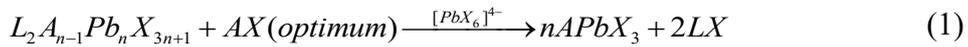

$$L_2A_{n-1}Pb_nX_{3n+1} + AX(optimum) \xrightarrow{[PbX_6]^{4-}} nAPbX_3 + 2LX \qquad (1)$$

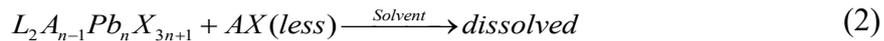

$$L_2A_{n-1}Pb_nX_{3n+1} + AX(less) \xrightarrow{Solvent} dissolved \qquad (2)$$

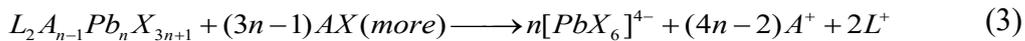

$$L_2A_{n-1}Pb_nX_{3n+1} + (3n-1)AX(more) \longrightarrow n[PbX_6]^{4-} + (4n-2)A^+ + 2L^+ \qquad (3)$$

In addition to the solution concentration, the type of ligand (L$^+$) in RP perovskites also impacts the conversion process by influencing the ion diffusion process as discussed in Supplementary Fig. 12 and Table S3. Considering both ion diffusion and structural rigidity, RP phase perovskites (n=3) with *n*-butylammonium (BA) ligand were found to be the optimal precursor for realizing the 2D APbX$_3$ structure. Other factors, such as annealing temperature (Supplementary Fig. 13-14), are also considered in our method (see details in Supporting Information). Through careful optimization, three kinds of 2D APbX$_3$ perovskites (Extended Data Fig.2), including MAPbI$_3$, MAPbBr$_3$ and FAPbI$_3$, were successfully realized. Particularly, uniform MAPbI$_3$ sheets with a variety of thicknesses and large lateral sizes were obtained by using a BA$_2$MA$_2$Pb$_3$I$_{10}$ perovskite as the template (Supplementary Fig. 15-18). These 2D crystals show good stability for nearly one month when stored in a nitrogen-filled glove box (Supplementary Fig. 19-22).

**Structure of 2D APbX$_3$ perovskites and their moiré superlattices**

The quality of 2D APbX$_3$ perovskites was systematically investigated. As shown in **Fig. 1b**, the typically acquired 2D MAPbI$_3$ perovskites with well-defined facets exhibit large single-domain size of over 10 μm on the arbitrary substrate (Supplementary Fig. 23), which represents an important advance over previously reported methods (Extended Table 1). Corresponding confocal photoluminescence (PL) mapping (**Fig. 1c**) indicates spatially uniform light emission from the converted sheets. Further, PL lifetimes are on the order of nanoseconds, which indicates low defect densities (Extended Data Fig. 3 and Table S4). High-resolution transmission electron microscopy (HRTEM) (Supplementary Fig. 24-27) was used to verify the single-crystalline nature and determine the in-plane crystal parameter *a* of 2D MAPbI$_3$. The HRTEM image (**Fig. 1d**) and selected area electron diffraction (SAED) pattern (**Fig. 1e**) both prove the four-fold symmetry in plane, and the in-plane crystal parameters are measured to be 0.63 nm for MAPbI$_3$. High-quality 2D MAPbBr$_3$ and FAPbI$_3$ perovskites were realized using a similar method and the details are summarized in Extended Data Fig. 4 and Supplementary Fig. 24-25.

As direct band gap semiconductors, 2D APbX$_3$ perovskites give strong PL emission, which is tunable by varying the small organic cations (A$^+$) and halide ions (X$^-$). The representative PL spectra are shown in **Fig. 1f** for ultrathin 2D MAPbBr$_3$, MAPbI$_3$, and FAPbI$_3$, whose emission peaks are blue-shifted from their bulk counterparts because of quantum and dielectric confinement (Supplementary Fig. 28-30). The emission linewidths are comparable to 2D RP phase perovskites with a high *n* number (Table S5), which indicates the high quality of our samples. The exciton binding energy in a 6-nm MAPbI$_3$ 2D sheet is estimated to be ~ 70 meV by fitting the absorption spectrum to Elliott's mode (Extended Data Fig. 3 and Supplementary Fig. 31), corresponding to an exciton radius of ~ 2 nm[44]. The exciton binding energy is much larger than that of 16 meV in the bulk[45-46] and it has been suggested that 2D and 3D perovskites converge when the layer thickness is greater than 13 nm[44]. MAPbI$_3$ exhibits the strongest thickness dependence in the PL emission peak position, which show the important role of interlayer interactions and the opportunities to modulate the energy landscape using these interactions.

Furthermore, X-ray diffraction (XRD) was used to further confirm the crystal structure of 2D ABX$_3$. Clearly shown in **Fig. 1g**, only two typical peaks, (001) and (002), are present for all three 2D APbX$_3$ perovskites, implying the complete conversion from RP phase perovskites and the lateral stacking along the [001] direction. In contrast to MAPbBr$_3$ and MAPbI$_3$, there is also a small shift of the reflection peaks to smaller diffraction angles for FAPbI$_3$, indicating its larger crystal parameters in alignment with the results in the reported literature[41]. The out-of-plane lattice parameter $c$ in 2D APbX$_3$ can be calculated from the XRD profiles, yielding a value of 0.60 nm for MAPbBr$_3$, 0.62 nm for MAPbI$_3$, and 0.63 nm for FAPbI$_3$. Together with TEM characterizations, a slight difference was found between $a$ and $c$, proving the quasi-cubic $Pm3m$ space group of the obtained 2D APbX$_3$ perovskite[47].

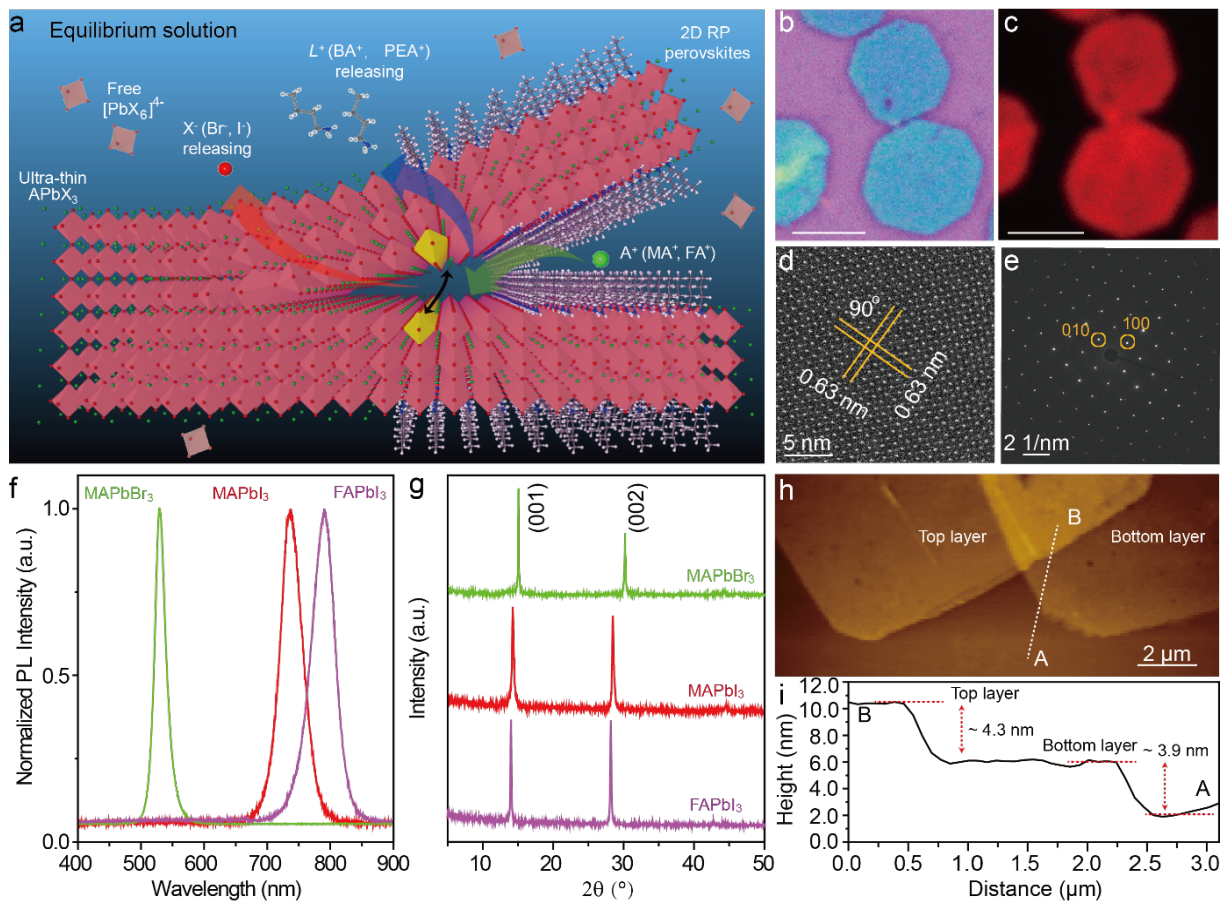

**Fig. 1| Conversion of RP phase 2D perovskite to APbX$_3$ phase via 'equilibrium solution' method and characterizations. a,** Schematic for the conversion mechanism from few layers RP perovskites to ultra-thin APbX$_3$ perovskites in equilibrium solution. **b,** Optical image of the typical ultra-thin MAPbI$_3$ perovskites obtained on SiO$_2$(300 nm)/Si substrate. The scale bar is 5 μm. **c,** The corresponding confocal

PL image of MAPbI$_3$ in **b** with the channel between 730 and 750 nm. The scale bar is 5 μm. **d**, High-resolution transmission electron microscopy (HRTEM) image of the typical single layer MAPbI$_3$ perovskite, indicating the equal lattice constants, *a* and *b*, along [001] zone axis. **e**, Corresponding SAED pattern of the single layer MAPbI$_3$ perovskite sheet, showing that it is single-crystalline. **f**, Steady-state PL emission spectra of three kinds of APbX$_3$ perovskites excited by 320 nm wavelength laser. **g**, XRD profiles of APbX$_3$ perovskite sheets obtained on silicon substrate with peaks indexed as (001) and (002). Particularly, for ultra-thin MAPbI$_3$ perovskites, the absence of splitting of the signals at 2θ=14.14° and 28.30° means a quasi-cubic phase[47]. It is possibly stabilized by substrate strain. **h**, AFM image of a twist-stacking APbX$_3$ perovskites. **i**, Corresponding height profile of the twist-stacking APbX$_3$ perovskites in **h**, showing a mean thickness of ~4 nm and nearly atomic flatness through roughness analysis.

Benefited from our facile synthesis method, direct mechanical stacking (see Supplementary Fig. 32 and Methods) was employed to construct artificial moiré superlattices with different twist angles. Briefly, a crystal was synthesized directly on polydimethylsiloxane (PDMS) and was subsequently transferred onto another ultrathin crystal on a silicon substrate (AFM image in **Fig. 1h**, also see optical images in Supplementary Fig. 33-34). The specific height profile in **Fig. 1i** shows the thickness of the 2D sheets to be around 4 nanometers (corresponding to ~6 unit cells [each unit cell ~0.62 nm thick])[41], and the surface is nearly atomically flat with a roughness (R$_a$) lower than 0.2 nm. More examples of individual sheets with a similar thickness and low surface roughness can be found in Extended Data Fig. 4. The 2D sheets were stacked along [001] zone axis and squared moiré patterns with the four-fold symmetry can be expected when the twisting angles are smaller than 20°, as illustrated in **Fig. 2a-c** and Supplementary Fig. 35. We then fabricated the TPLs on ultrathin SiN$_x$-based TEM grids and characterized them using HRTEM. As shown in **Fig. 2d-f**, the square moiré patterns were clearly visualized. The periodic lengths of different square moiré patterns are measured as 13.0, 5.1, and 2.3 nm for twist angles of 2.5°, 7.3° and 15.0°, respectively (**Fig. 2d-f**). These are further confirmed by the simulations under different imaging modes, i.e., TEM and scanning transmission electron microscopy (STEM) modes (Supplementary Fig. 36-39 and Extended Data Fig. 5). SAED patterns of the different twisted samples (insets of **Fig. 2d-f**) clearly show two sets of diffraction spots giving the exact twist angle.

Additionally, using HRTEM, we were able to determine the crystal orientation from the edges of the 2D sheets with well-defined facets (see Supplementary Fig. 40 for detailed discussions). This information is important because it later allows us to determine the exact twist angle simply from optical microscope images by measuring the intersection angle between specific crystal edges without the need for measuring HRTEM images on every sample (Supplementary Fig. 34).

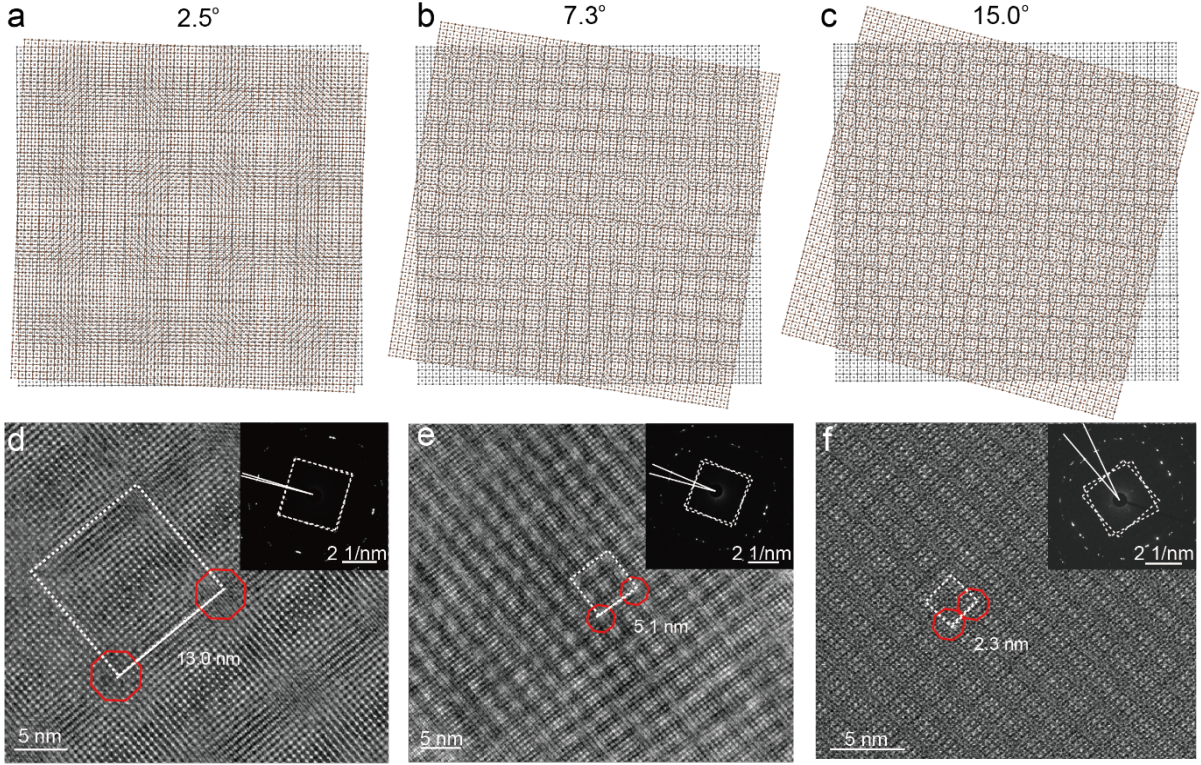

**Fig. 2| Square moiré patterns in twisted perovskite layers (TPLs). a, b, c,** Structure models of twisted bilayers of MAPbX$_3$ with twist angles of 2.5° (**a**), 7.3° (**b**) and 15.0° (**c**), using the simplified lattice and stacked along [001]. **d, e, f,** HRTEM images of the TPLs with different twist angles, including 2.5° twisted MAPbBr$_3$ (**d**), 7.3° twisted MAPbI$_3$ (**e**) and 15.0° twisted MAPbBr$_3$ (**f**), presenting the clear square moiré patterns and different periodic lengths.

### Exciton transport in perovskite moiré superlattice

The formation of moiré superlattice has been demonstrated to impose a spatially modulating potential (moiré potential) on a single layer of 2D materials. Earlier theoretical work predicted the TPLs stacked by MA$_2$PbI$_4$ can yield flat bands that are expected to have profound effects on exciton properties[14] but are yet to be investigated experimentally. Here, we first employed time-resolved PL microscopy (details in Methods section and Supplementary Fig. 41) to measure exciton migration as a function of twist angle in TPLs. This technique provides spatial maps of excitons with a time resolution of ~100 ps[50-52]. The emission from the excitons was projected by an imaging lens onto a time-correlated single-photon counting detector which was scanned to provide temporally and spatially resolved PL profiles. The transport measurements were performed at an exciton density of 2×10$^{10}$ cm$^{-2}$, well below the onset of exciton–exciton annihilation (EEA).

We compared exciton diffusion in MAPbI$_3$ TPLs with different twist angles, and results from 2°, 9° and 29° are shown in **Fig. 3a-c** and Supplementary Fig. 42. The time- and space-dependent exciton density is given by, $\frac{\partial N(x,t)}{\partial t} = D\frac{\partial N^2(x,t)}{\partial x^2} - \frac{N(x,t)}{\tau}$, where $D$ is the diffusion constant, and $\tau$ is the exciton lifetime. The exciton density $N(x, t)$ is proportional to the PL counts and exciton migration led to a broader Gaussian distribution at a delayed time $t$, $N(x,t) = N_t \exp(-\frac{x^2}{2\sigma_t^2})$ where $\sigma_t^2$ is the variance (**Fig. 3d-f**). The experimentally measured $\sigma_t^2$ - $\sigma_0^2$ corresponds to the mean squared distance travelled by the excitons, and the diffusion constant is given by $D = (\sigma_t^2 - \sigma_0^2)/2t$ (**Fig. 3g**). Remarkably, exciton diffusion is strongly dependent on the twist angle, which provides experimental evidence for the moiré potential in TPLs. For instance, exciton motion in TPLs with a 9° twist angle is significantly impeded, where $D$ decreases by a factor of 2-3 in the junction region (**Fig. 3g**). The statistics of $D$ measured for 16 total TPLs further confirm the localization of excitons at twist angle around 10° (Extended Data Fig. 6d and Table S6). While, for a large twist angle of 29°, the $D$ values of the TPLs lie in between that of the top and bottom layers. At this large angle, the moiré period is smaller than the intrinsic lattice spacing and thus the effect of moiré potential should be negligible. Interestingly, for TPLs with a 2° twist angle, $D$ increases by a factor of 3 (Extended Data Fig. 6d) compared to the individual layers, opposite from the TPLs with a 9° twist angle. **Fig. 3h** schematically illustrates the exciton localization in TPLs due to the imposed moiré potential. The fact that exciton localization observed in TPLs constructed from slabs of up to 9-unit-cell thick (~ 5 nm) indicates much longer-range interlayer interaction than in twisted graphene and TMDCs.

Twist-angle-dependent exciton dynamics also provides additional supports for the existence of localized excitons. With the mobile excitons, the PL decay of individual layers and the junctions in TPLs with small and large twist angles are strongly pump intensity-dependent. As the pump intensity increases, a faster decay component emerges (**Fig. 3i** and Supplementary Fig. 43). These can be interpreted by EEA, where two excitons interact with each other and fuse to form a higher energy exciton followed by rapid internal conversion to the lowest energy excited state (Supplementary Fig. 44). For the 2° and 29° twist angles, EEA rates were extracted to be $2.2 \times 10^{-3}$ cm$^{-2}$s$^{-1}$ to $2.1 \times 10^{-3}$ cm$^{-2}$s$^{-1}$, respectively (**Fig. 3i**), which are similar to the rates in individual layers (Extended Table 2 and Supplementary Fig. 43) and previously reported values in 2D perovskites[53]. The variations in the EEA rates in the individual layers are likely due to differences in thickness. In contrast, for the TPLs with a twist angle near 9°, such an excitation density dependence in recombination was negligible (**Fig. 3i**) and the EEA rate, 4.5 $\times 10^{-5}$ cm$^{-2}$s$^{-1}$, was two orders of magnitude lower than that in the top and bottom layers (Extended Table 2). The twist-angle dependence of EEA further reinforces the notion of exciton localization by the moiré potential in the case of a 9° twist angle. This localization effect results in a decreased likelihood of exciton encounters, leading to a reduced rate of EEA. Note that the highest exciton density investigated here is on the order of ~ $10^{11}$ cm$^{-2}$, below one exciton per moiré site.

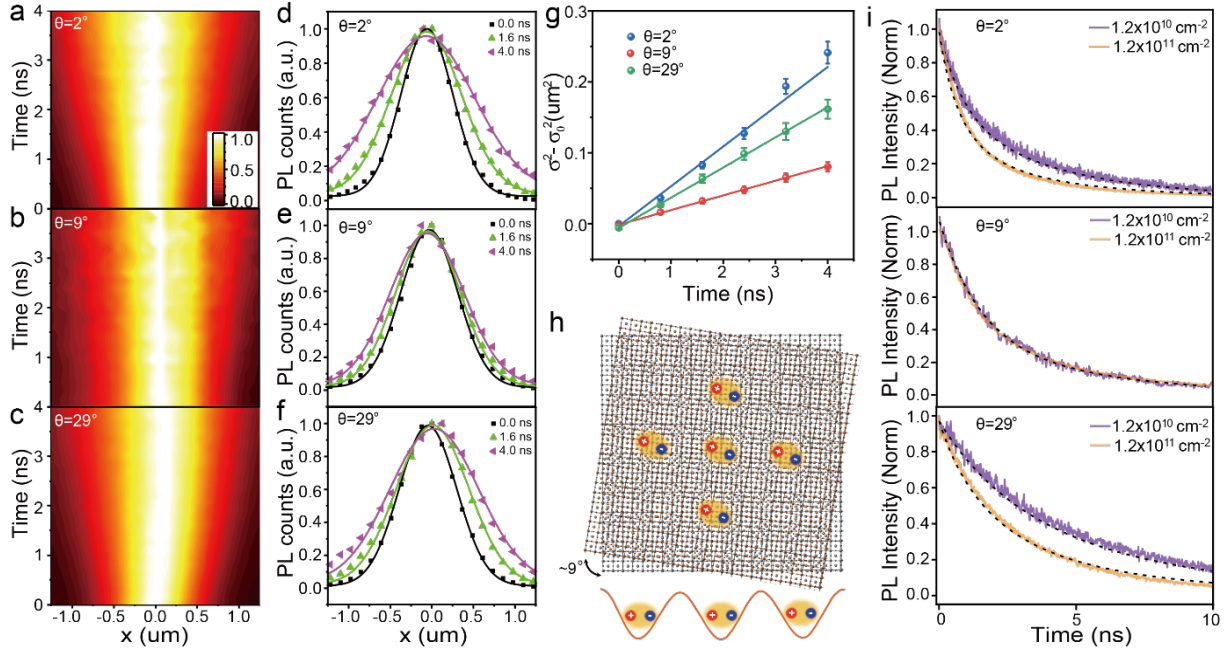

**Fig. 3| Twist angle-dependent exciton transport and annihilation in MAPbI₃ TPLs. a-c,** Time-resolved PL imaging of exciton diffusion in MAPbI₃ TPLs with different twist angles under different pump-probe delay times at 295K, ~2° (**a**), ~9° (**b**) and ~29° (**c**). Slower exciton diffusion can be observed in the MAPbI₃ TPLs with a twist angle of ~9°. **d-f,** Exciton density distribution at representative delay times in the TPLs with different twist angles of ~2° (**d**), ~9° (**c**) and ~29° (**f**). Data are extracted from part **a-c**. Solid lines are fits using a Gaussian function to obtain $\sigma_t^2$. **g,** Determining the corresponding exciton diffusion coefficient by fitting $\sigma_t^2$ as a function of time in TPLs with different twist angles of ~2°, ~9°and ~29°. The error bars indicate the standard error of Gaussian fit of the exciton density distribution profiles. **h,** Schematic illustration of exciton localization by moiré potentials (bottom) in MAPbI₃ TPLs (top) with a twist angle of ~9°. **i,** Time-resolved photoluminescence (TRPL) decay curves with different exciton density for MAPbI₃ TPLs having twisting angles of ~2°, ~9° and ~29°. Exciton-exciton annihilation rate is reduced by two orders of magnitudes in the 9°-twisted TPLs because of the localized excitons. Dashed lines are fits to the model described in the Method.

Further, we fabricated two-terminal devices to examine their charge transport characteristics at room temperature (Extended Data Fig. 7). The TPLs with twisting angles of 9.0° exhibit an electrical conductivity of ~1.2×10⁻⁵ S/m, significantly lower than that of the twisting angles of 1.5° (~3.1×10⁻² S/m) and individual layer (~2.9×10⁻⁴ S/m). Assuming the carrier concentrations are similar in these samples (as they are synthesized under the same condition), these results suggest the lowest charge carrier mobility in the 9.0° sample. This is consistent with the observation in exciton diffusion measurements discussed above, as well as the increase in electrical conductivity at lower twisting angles. The similar trend again reflects the strongly correlated electronic phases modulated by moiré potentials.

**Twist-angle-dependent bright exciton emission and absorption**

We next examine the light emission and absorption properties of these TPLs. As shown in **Fig. 4a-c** and Supplementary Fig. 45, the PL emission intensity and reflectance contrast were found to be strongly dependent on the twist angle. The confocal PL mapping indicates that the twist-angle-dependent PL enhancement is spatially uniform, which excludes spatial heterogeneity effects. With an excitation beam size of ~ 0.4 μm, we can reliably record PL spectra separately from the junction as well as the top and bottom layer-only areas (Supplementary Fig. 33). The PL enhancement can also be seen in the twist-angle-dependent emission spectrum shown in **Fig. 4d-f** and Supplementary Fig. 46-48.

Remarkably, detailed statistics implies that the PL intensity was enhanced by more than one order of magnitude in the twist angle range of 8-12°, as summarized in **Fig. 4g** and Table S7. In addition, the reflectance contrast was also enhanced in the junction area over the individual layers at the twist angle of ~ 10° (Extended Data Fig. 8), which can be explained by the enhanced imaginary part of the refractive index, as shown in Supplementary Fig. 49. The latter is directly proportional to the exciton absorption, which is associated with the oscillator strength. The striking enhancement in PL emission at ~ 10° is likely the result of the localized bright moiré excitons in TPLs, highlighting a crucial distinction from TMDCs. Although moiré potentials have been observed in twisted TMDC bilayers, optically dark momentum-indirect K-Q excitons form as a consequence of the interlayer interactions predominantly occurring at the Q point of the Brillouin zone[55,56]. We also examined the temperature dependence of the PL enhancement (**Fig. 4h** and Supplementary Fig. 50-51). The enhancement factor increases from 5.6 at room temperature to over 20 at 160 K, which is consistent with the localized moiré excitons being a bright state and the ground state of the system. We also inserted a hexagonal boron nitride (hBN) layer to disrupt interlayer coupling in a control sample with a 10.5° twist angle (Supplementary Fig. 48). This sample shows no PL enhancement, which further substantiates that the observed enhancement is indeed a result of moiré effects.

We note that the PL intensity was enhanced but to a less extent for the twist angles smaller than 5°. At large twist angles (> 20°), the PL intensity (**Fig. 4g**) and reflectance contrast (Extended Data Fig. 7) at the junction were similar to the sum of the two layers (no obvious enhancement). Interfacial defect passivation likely plays a role in smaller twist angles TPLs due to improved lattice matching. As a result, it may account for the slight PL enhancement (as well as the exciton and charge transport enhancement discussed earlier) in TPLs with smaller twist angles (<5°). Nevertheless, it is important to stress structural defects are not the dominating factor for the observed exciton localization and light emission/absorption enhancements at ~10°. For instance, if defects are the reason for exciton localization, PL should be significantly quenched, rather than dramatically enhanced.

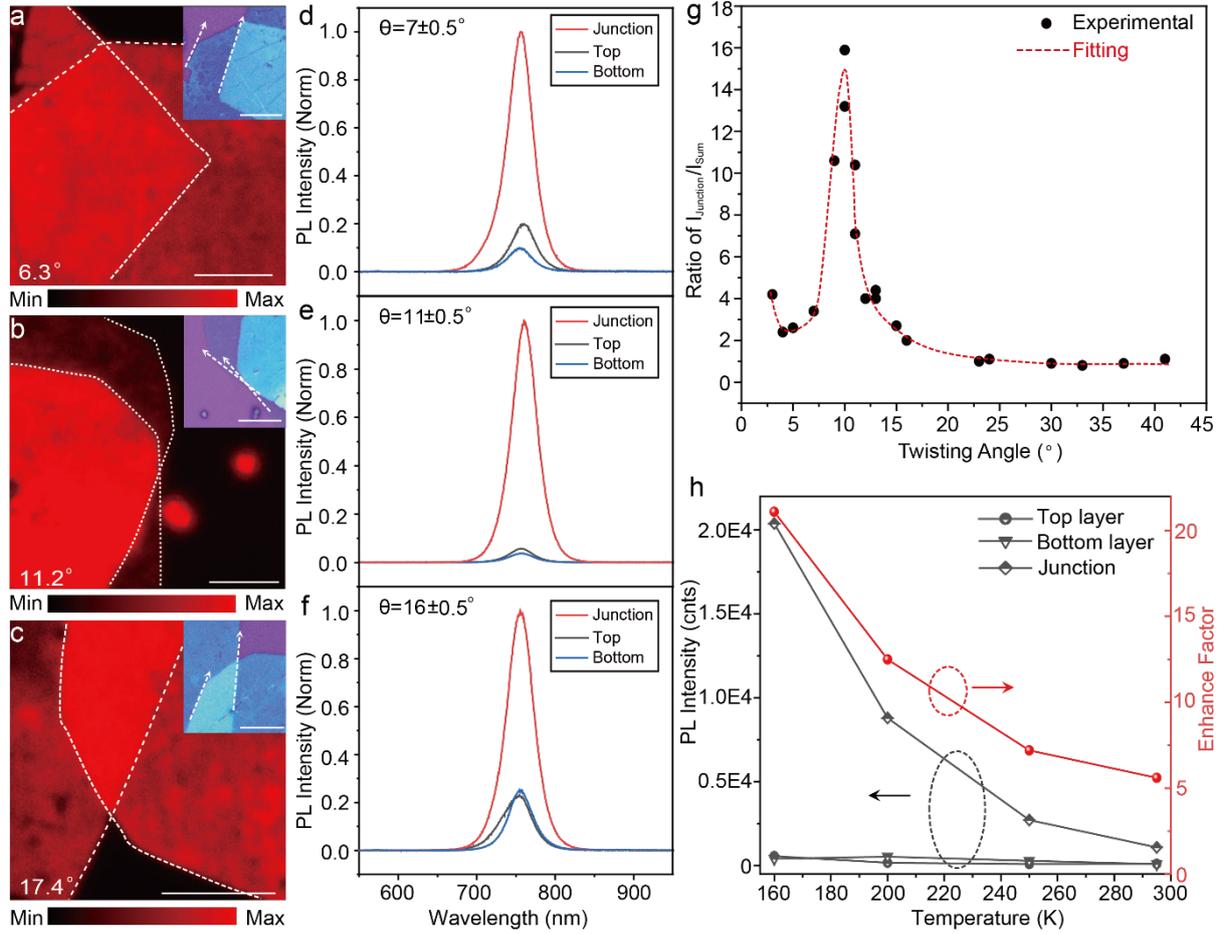

**Fig. 4| Twist angle-dependent PL emission in MAPbI₃ TPLs. a-c,** Spatially-resolved confocal PL mappings of the MAPbI₃ TPLs with different twist angles using a 405 nm laser as the excitation wavelength and emission channel widths of 730-755 nm, ~6.3° (**a**), ~11.2° (**b**), and ~17.4° (**c**). The insets are the corresponding optical images of MAPbI₃ TPLs. The dashed line is used to mark the edge of samples to show the intersection angle with a twist. All the scale bars are 10 µm. These results show that the enhancement of the PL is spatially homogeneous. **d-f,** Comparison of steady-state PL emission spectrum in MAPbI₃ TPLs with different twist angles of ~7° (**d**), ~11° (**e**) and 16° (**f**), indicating more than one order of magnitude enhancement of PL emission intensity at the junction region in TPLs of ~11°. **g,** Summary of angle-dependent PL emission in TPLs by acquiring the ratio of PL intensity at junction part to sum of two individual layers' intensity, showing the great intensity enhancement for the TPLs with twist angles of 8~12°. The dashed line is a fitting to show the trend on the changes of PL intensity. **h,** Temperature-dependent PL intensities for individual layers and junction in the TPLs with a twist angle near 10°, and PL enhance factors before phase transition.

## Theoretical understanding of moiré flat bands in TPLs

To gain insights into the electronic structure of TPLs, we performed first-principles calculations based on time-(in)dependent density functional theory (DFT) (**Fig. 5a-f** and Supplementary Fig. 52-56). The computational model for a twisted MAPbI₃ bilayer contains two halved crystals, each with two inorganic layers and are twisted relative to each other by an angle $\theta$ (Supplementary Fig. 53). For $\theta=9.5°$, the moiré supercell has 2409 atoms with a negligible lattice mismatch (Supplementary Fig. 53); this twist angle is chosen because it is computationally tractable and is close to 10° as examined in the experiment. The single-particle band structure for the TPLs ($\theta=9.5°$) is displayed in **Fig. 5a** along with its relaxed atomic structure at 0 K (**Fig. 5b**) and the lowest exciton charge density (**Fig. 5c**). It is found the I-Pb bonds at the interface are bent (shown in red rectangles) due to the strong I-I repulsion across the interface. The bending only occurs at the AA stacking where the interlayer distance is the shortest. The bent I-Pb bonds yield a relatively large bandwidth (~120 meV for the valence band maximum, VBM) and delocalized hole charge density (cyan) across the top layer.

However, if the bent I-Pb bonds are straightened (**Fig. 5e**), remarkable changes to the band structure are observed; the VBM becomes very flat with a bandwidth of ~10 meV (**Fig. 5d**), accompanied by strong localization of the hole (**Fig. 5f** and Extended Data Fig. 9). In particular, the hole density becomes connected across the interface thanks to the straightened I-Pb bonds that reduce the interfacial I-I distances and enhance the interlayer interactions (Extended Data Fig. 9). The localized hole (or VBM state) originates from the interlayer coupling at the AA stacking, consistent with the previous work[14]. Straightening the bent I-Pb bonds costs ~3.9 meV per atom, which is much less than the thermal energy at 300 K (~39 meV per atom). This implies that the thermal fluctuation could unbend the I-Pb bonds and lead to a flat VBM in the TPL with $\theta=9.5°$ at the room temperature. One distinctive feature of hybrid perovskites is their soft and dynamic nature. A static structure at 0 K merely represents a snapshot among a myriad of possible structures. Thus, we contend that even minor distortions from the 0 K structure can offer critical insights into the formation of flat bands. As a control, we have carried out two single-point calculations to obtain the band structure for each layer, as shown in Supplementary Fig. SX. Both individual layers have larger bandwidth than the twisted bilayer. This further supports that the flat band results from the moiré effects.

We also carried out the similar calculations for a TPL with $\theta=16.5°$. With a larger twist angle, the average interlayer distance near the AA stacking is increased, and the bending of the I-Pb bonds is negligible; the VBM bandwidth is found to be ~70 meV (Supplementary Fig. 54). In contrast, the VBM bandwidth for a non-twisted MAPbI$_3$ bilayer (AA stacking) is 510 meV (Supplementary Fig. 55). Unfortunately, DFT calculations for a TPL structure with $\theta < 9.5°$ are beyond our computational capacity. Nevertheless, it is reasonable to believe that there exists a "magic" angle or angles between 9.5° and 16.5° (much closer to the former than the latter) which could host flat bands and moiré excitons. At such an angle, the interlayer distances are optimal so that the Pb-I bonds are nearly straight while the I-I distances are short enough. In other words, the DFT calculations provide a strong support of the experimental observation of a special angle ~10°.

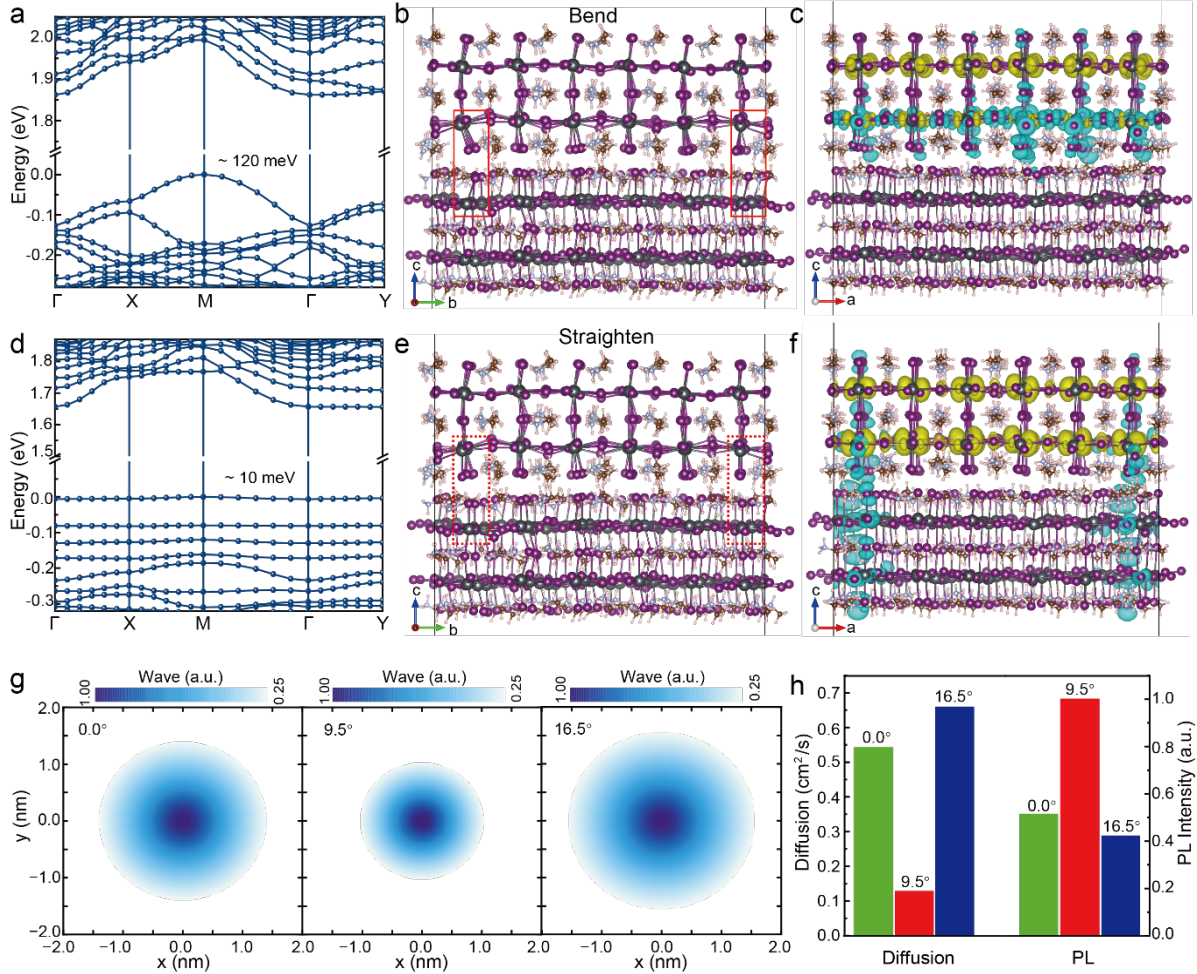

**Fig. 5| Theoretical calculations of the electronic structure and excitonic properties in MAPbI₃ TPLs (θ =9.5°). a, d,** The single-particle band structure of the TPL in the equilibrium structure at 0 K (**a**) and in a modified structure with straightened I-Pb bonds (**d**). The Fermi level is set to zero. **b, e,** Side-view of the TPL in the equilibrium structure (**b**) and the modified structure with straightened I-Pb bonds (**e**). The solid red rectangle highlights the bent I-Pb bonds and the dashed rectangle indicates the straightened I-Pb bonds. **c, f,** The charge density of the lowest energy exciton in the equilibrium structure of TPL at 0 K (**c**) and in the modified structure with straightened I-Pb bonds (**f**). The hole and electron densities (iso-surface value set at 0.0005 eÅ⁻³) are shown in cyan and yellow color, respectively. **g,** The excitonic wavefunctions in real space for TPLs with different twist angles, 0.0° (left), 9.5° (middle) and 16.5° (right), by only considering 1s exciton energy level. The wavefunctions have been obtained by solving the Wannier equation for the investigated material. **h,** Theoretical calculations on angle-dependent PL emission and diffusion properties of MAPbI₃ TPLs with the consideration of 1s excitons as the main contribution.

We further computed the twist-angle-dependent 1s exciton wavefunctions using the Wannier equation[48-49], taking into account the effective mass of electrons and holes acquired from the above electronic structure calculations. The exciton size is clearly angle-dependent, with the most strongly bound exciton exhibiting the smallest radius and heaviest mass in the MAPbI$_3$ TPLs at a twist angle of 9.5° (**Fig. 5g**). The enhancement of the effective mass, as a function of angle, directly leads to a reduction in the diffusion constant, due to the inverse proportionality, as illustrated in **Fig. 5h**. Meanwhile, the oscillator strength of 1s excitons based on a microscopic many-particle model[54] and density of states (Supplementary Fig. 56) were calculated and demonstrate an almost 3-fold increase in PL intensity for 9.5° compared to 16.5°, consistent with the experimental results, as depicted in **Fig. 5h**. Overall, due to the proper interface distance for I-I coupling and soft ionic lattice, moiré flat bands appeared in the TPLs with a twist angle of near 10° and induces the localized exciton/carrier, described by Supplementary Fig. 57.

**Discussion and outlook**

Our results demonstrated twisted 2D APbX$_3$ perovskite moiré superlattices as a new platform for moiré flat bands beyond vdW interaction. Remarkably, the ionic interlayer interaction is much longer-range than conventional 2D vdW materials (usually limited to 1- or 2-unit-cell thickness)[21], while our 2D perovskites are up to 9-unit-cell thick. The strong ionic interactions also allow flat band effects to be observed at higher temperatures, making these perovskite structures a new class of room-temperature moiré materials. Distinct from the dark excitons in twisted TMDC bilayers, the bright direct moiré excitons in halide perovskites are particularly exciting for the exploration of light emission and light-matter interactions, such as lasers and formation of exciton polaritons[57-59]. The enhanced oscillator strengths of the excitons also provide new opportunities to design new energy and charge transfer functionalities[60]. Our findings offer design principles for thin film perovskite optoelectronic devices through the manipulation of twist angles between adjacent layers. Specifically, employing small twist angles to promote enhanced charge transport can be beneficial for solar cells, whereas utilizing a "magic" twist angle to enhance light emission can be advantageous for LEDs.

A key advantage of halide perovskites is its highly programmable structures. As a demonstration of structural tunability, we have performed measurements on twisted FAPbI$_3$ layers (shown in Supplementary Fig. 58-59 and Table S8). Similar PL enhancement was observed but to a less extent than the twisted MAPbI$_3$ layers. This indicates the interlayer coupling can be tuned using different cations. Further investigations of the lattice relaxation effects on the moiré potential are also warranted because the easily deformable metal halide bonds can potentially make the flat bands even more robust against defects and fluctuations. Our calculations have hinted these effects to some extent, as the relaxed structure shows the delayed appearance of the moiré flat bands. In future work, moiré patterns can also be designed to exploit unique properties of halide perovskites, such as Rashba splitting and ferroelectricity.

Finally, moiré patterns in square perovskite lattices have been theoretically investigated decades ago and many new physical phenomena have been predicted, including high-temperature superconductivity[61], multiferroicity[62], and piezoelectricity[63]. Unfortunately, these structures have not been realized experimentally and their properties remain to be explored. Our work provides a first example of square moiré superlattice in perovskites, which paves the way for future experimental realization of a wider range of materials, including but not limited to halide perovskites, oxide perovskites, and a variety of chalcogenide compounds, toward novel optoelectronic, magnetic, superconducting, and topological properties. The structural tunability, and unique materials properties in combination with bonding beyond vdW interaction open exciting new areas of moiré materials.

## Online content

Any methods, additional references, Nature Research reporting summaries, source data, extended data, supplementary information, acknowledgements, peer review information; details of author contributions and competing interests; and statements of data and code availability are available at XXX.

## Methods

**Chemicals and reagents.** Organic solvents, including anhydrous chlorobenzene (CB), isopropyl alcohol (IPA) and solid chemicals, including lead bromide (PbBr$_2$) and lead iodide (PbI$_2$) were purchased commercially (Sigma Aldrich). n-butylammonium bromide (BA·HBr), n-butylammonium iodide (BA·HI), phenethylammonium iodide (PEA·HI), methylammonium bromide (MABr) and methylammonium iodide (MAI) were purchased commercially (Greatcell Solar). All above chemicals were used as received. Other ammonium salt ligands, such as bithiophenylethylammonium iodide (2T·HI), were synthesized in our lab.

**Synthesis of 2D halide perovskite crystals.** The synthesis of 2D RP phase halide perovskite crystals is based on the following recipes. For BA$_2$PbI$_4$ and PEA$_2$PbI$_4$ crystals, PbO (0.57 mmol) and BAI or PEAI (0.57 mmol) precursors are dissolved into an acid mixture containing 0.9 mL of HI and 0.1 mL of H$_3$PO$_2$ in a 10 mL glass vial. For BA$_2$MAPb$_2$I$_7$ crystal, PbO (0.59 mmol), BAI (0.43 mmol), and MAI (0.31 mmol) precursors are dissolved into an acid mixture containing 0.9 mL of HI and 0.1 mL of H$_3$PO$_2$ in a 10 mL glass vial. For BA$_2$MA$_2$Pb$_3$I$_{10}$ crystal, PbO (0.59 mmol), BAI (0.19 mmol), and MAI (0.4 mmol) precursors are dissolved into an acid mixture containing 0.9 mL of HI and 0.1 mL of H$_3$PO$_2$ in a 10 mL glass vial. For all the perovskite crystals synthesis, with magnet stirring, the vial is heated to 120 °C in an oil bath. After the solid precursors are completely dissolved and the solution becomes transparent, the stirring is terminated, and the solution is cooled at a cooling rate of 10 deg/5 min, during which the crystals are formed. The final crystals are collected by vacuum filtration, and the residue solvent is removed via vacuum pumping.

But for $(2T)_2MA_2Pb_3I_{10}$ crystal growth, 500 µL isopropyl alcohol (IPA) is added in the mixture of 400 µL hydroiodic acid (HI) and 200 µL hypo phosphorous acid ($H_3PO_2$) to assist the crystallization process and dissolve conjugated ligand precursors (5mg 2T·HI). Also, in the solution, 40 mg $PbI_2$ and 50 mg MAI are used as the precursors. The vial is heated in a water bath until all precursors were dissolved typically within 5 mins and naturally is cooled down to room temperature to get the single crystal.

**Synthesis of quasi-2D halide perovskite nanocrystal.** The solution-air interface method similar to that reported[34] is adopted to obtain quasi-2D halide perovskite nanocrystals. Specifically, PbO (0.59 mmol), BABr or BAI (0.19 mmol), and MABr or MAI (0.4 mmol) precursors are dissolved into an acid mixture containing 0.9 mL of HBr or HI and 0.1 mL of $H_3PO_2$ in a 10 mL glass vial. With magnet stirring, the vial is heated to 120 °C in an oil bath. The $BA_2MA_2Pb_3Br_{10}$ solution was cooled down to 28°C, while $BA_2MA_2Pb_3I_{10}$ solution was cooled down to 38°C, and kept at that temperature in a closed vial as the stock solution. Two microliters of this warm supernatant solution were collected and dispensed with a 10 µL pipette onto a glass slide placed in an open ambient environment. On the solution surface, $BA_2MA_2Pb_3Br_{10}$ or $BA_2MA_2Pb_3I_{10}$ nanocrystals could be got and picked up with the help of PDMS.

**Solution conversion to synthesize 2D ultra-thin $APbX_3$ perovskites.** A series of $APbX_3$ equilibrium IPA solutions, in which the mole ratio of $AX/PbX_2$ is 3/1, was made to realize the conversion to ultra-thin $APbX_3$ nanocrystals. Specifically, the concentration of AX for $MAPbI_3$, $MAPbBr_3$, and $FAPbI_3$ is10 mg/mL, 12 mg/mL and 11.2 mg/mL, respectively. Before use, the equilibrium solution was stored at 45°C for 12h to reach solid-liquid reaction equilibrium between $PbX_2$ solid and MAX (X=Br, I). For exfoliated 2D RP phase perovskite samples, to realize the complete conversion, 1mL $APbX_3$ equilibrium IPA solution is used and the reaction lasted for 30 min. After the reaction, 1mL mixed CB/IPA (v/v=1mL/50 µL) solution was used to wash samples, which should be annealed further. Different annealing temperature and time are adopted, 120 °C and 8 mins for conversion to $MAPbBr_3$, 55 °C and 10 mins for conversion to $MAPbI_3$ and $FAPbI_3$. When using 2D RP phase perovskite nanocrystals as precursors, 240 µL $APbX_3$ equilibrium IPA solution is used for samples on the silicon substrate and 260 µL for samples on PDMS. For conversion to $MAPbBr_3$ and $MAPbI_3$, the reaction time is 15~20 mins, but for conversion to $FAPbI_3$, only 5~10 mins are enough. The annealing temperature and time are the same as using exfoliated samples.

**Fabrication of perovskite twist structure.** For the typical $MAPbI_3$ TPLs, ultra-thin $MAPbI_3$ on PDMS and silicon substrates were obtained separately and vertical junction was realized by point to point transfer using the equipment shown in Supplementary Fig. 32. Based on this, twist angles in vertical junctions could be controlled. For $FAPbI_3$ TPLs (Supplementary Fig.

58) and MAPbBr$_3$ TPLs (Supplementary Fig. 59), similar method was used.

**Characterizations**

**Optical imaging.** The bright-field optical images were collected by a custom microscope (Olympus BX53).

**Photoluminescence imaging and spectra collection.** Samples were excited with a light source (012-63000; X-CITE 120 REPL LAMP). The filter cube contains a bandpass filter (330–385 nm) for excitation, and a dichroic mirror (cutoff wavelength, 400 nm) for light splitting and a filter (long pass 420 nm) for emission. The photoluminescence spectra were collected by a spectrometer (SpectraPro HRS-300).

**XRD measurements.** XRD was measured using a powder X-ray diffractometer (Panalytical Empyrean) with a Cu Kα source. The wavelength (λ) is 0.154 nm. The XRD measurements were performed on the as-grown ultra-thin APbX$_3$ perovskites covering almost the entire surface of the SiO$_2$ (300 nm)/Si substrate, which were realized by being converted from 2D RP phase perovskites exfoliated on the substrate.

**AFM.** AFM images were recorded in tapping mode using an atomic force microscope (Bruker MultiMode 8).

**IR characterization.** Attenuated total reflectance (ATR) – Fourier transform infrared (FTIR) spectroscopy was measured with Thermo Nicolet Nexus FT-IR, equipped with MCT detector, KBr beam splitter, and diamond ATR. To collect signal with enough intensity, the micro-sized 2D RP phase perovskites (n=1) with PEA$^+$ ligands were used for the conversion. And the samples for FTIR measurement were dispersed on SiO$_2$(300 nm)/Si substrate, and the measurements were performed with the wavenumber range of 800-4500 cm$^{-1}$.

**Reflectance, steady-state and time-resolved PL**

Reflectance, steady-state and time-resolved PL measurements were carried out using a 40x [numerical aperture (NA), 0.60] objective in a home-built microscope setup. For reflectance or PL measurements, a white light source (Thorlabs) or a laser light generated by a picosecond-pulsed diode (LDH-P-C-450B, PicoQuant) with a 2.8 eV excitation energy (full width at half maximum, 50 ps) was focused by the objective. For power density-dependent time-resolved PL measurements, the power, 20 nW, 30 nW, 50 nW, 90 nW and 200 nW, are used and converted into the exciton density as:

$$n = \frac{\alpha(\lambda) * P_f}{h\nu}$$

where α (λ), $h\nu$ and $P_f$ are the absorption, photon energy (447 nm) of pump pulse and the peak fluence of pump pulse respectively. The peak fluence of pump pulse ($P_f$) could be calculated as:

$$P_f = \frac{P}{A * f}$$

where $P$ is the average laser power, $A$ is the effective area of the pump beam (the diameter is 450 nm), $f$ is the repetition rate of the laser ($4 \times 10^7$ Hz).

At the same time, the absorption ($\alpha$ ($\lambda$)) is calculated as:

$$\delta R(\lambda) = \frac{4}{n_{Sub}^2 - 1} \alpha(\lambda)$$

where $\delta R(\lambda)$ is the differential reflection and $n_{Sub}$ is the refractive index of the SiO$_2$ (300 nm)/Si substrate. In our experiments, the differential reflection could be got in Extended Data Fig. 7, and it is 0.94 for junction and 0.90 for individual layer. $n_{Sub}$ is 1.459.

The reflected white light or PL emission was collected with the same objective, dispersed with a monochromator (Andor Technology), and detected by a thermoelectric cooled charge-coupled device (Andor Technology). The reflectance difference spectra were derived by calculating the difference between the reflected white light spectra from the sample and the substrate, $\Delta R(\lambda) = R(\lambda) - R_0(\lambda)$, where R is the reflected white light intensity from the sample and $R_0$ is the reflected white light intensity from the substrate. Time-resolved PL measurements were performed using a time-correlated single-photon counting module (PicoQuant) with a time resolution of ∼ 100 ps.

Temperature-dependent steady-state PL measurements were performed in a helium cooled cryo-station (MONTANA INSTRUMENT), from 13 K to 298 K. This experiment indicates an occurrence of phase transition, from quasi-cubic (or pseudo-cubic) to orthorhombic, at 160-120 K (Supplementary Fig. 50-51), and thus impedes further exploration on flat band at low temperature. A ×40 (Nikon, NA = 0.60) objective was used to focus onto the sample.

To acquire the credible data, optical measurements on individual layer and junction part were done at the same time, because the exposed each parts are much larger than the laser beam size (Supplementary Fig. 33-34).

**Exciton-exciton annihilation (EEA) rate of individual layer and MAPbI$_3$ TPLs**

The power-dependent TRPL measurements were performed with the individual layer MAPbI$_3$ as well as the homojunctions of different twist angles using the same equipment with normal time-resolved PL measurements. The exciton decay kinetics were fitted using the rate equation below:

$$\frac{\partial n_e(x,t)}{\partial t} = -k_e n_e(t) - 2 * k_{EEh} n_e(t)^2,$$

where $k_e$ is the exciton monomolecular decay rate and $k_{EEA}$ is the EEA rate. The extracted $k_{EEA}$ is in the range of $1e^{-3}$ to $3e^{-3}$ $cm^2s^{-1}$ for the individual few-layer $MAPbI_3$, and it is comparable to $(BA)_2(MA)_{n-1}Pb_nI_{3n+1}$ 2D perovskites[53]. For the $MAPbI_3$ TPLs with the twist angle near 10°, $k_{EEA}$ is two orders of magnitude smaller. For the homojunctions with large or small twist angles, $k_{EEA}$ is comparable to the individual $MAPbI_3$.

**Fluorescence-lifetime imaging microscopy measurements**

A home-built PL microscope was used to conduct time-resolved PL imaging (TRPLI) measurements. A picosecond pulsed laser with a wavelength of 447 nm was used as the excitation source (10 MHz repetition rate). The beam was focused onto the sample by using a long working distance 40× objective (Nikon, NA = 0.6). The epic scattered PL emission was collected by the same objective, and guided and focused onto a single photon avalanche diode (PicoQuant, PDM series) with a single photon counting module (PicoQuant). To elucidate exciton dynamics in individual layer and in junction region, power-dependent TRPL was performed with a motorized continuously variable ND filter (Thorlabs) to control the excitation density.

For PL imaging measurement, a 50× objective (Olympus, NA = 0.95) was used to achieve a better spatial resolution. A 600 nm imaging lens was used in the light path which results in an overall magnification of 167×. By scanning the avalanche diode at the image plane with a motorized stage (Thorlabs), time-resolved PL imaging around the excitation spot was acquired.

**Confocal PL imaging**

A Leica SP8 inverted laser scanning confocal microscope with a 405 nm laser line as the excitation wavelength was used for the PL mapping of the TPLs on $SiO_2$ (300 nm)/Si substrate.

**TEM characterizations**

**TEM sample preparation of pure 2D $APbX_3$ nanocrystals.** To obtain $APbX_3$ perovskite nanocrystal on TEM grids, two strategies were adopted. One is realizing the direct conversion to $APbX_3$ perovskites on copper grid with 1500 mesh by directly taking 2D RP phase perovskite nanocrystal from solution on it. The other one is transferring the ultra-thin $APbX_3$ on PDMS to the hole of $SiN_x$ grids. In the pore, there is a 10 nm $SiN_x$ amorphous film as the support. Two kinds of samples were both used to characterize the $APbX_3$ perovskite structure in these experiments.

**TEM sample preparation of 2D $APbX_3$ perovskites vertical junctions with different twist**

**angles.** Only adopting $SiN_x$ grid, on which there is a pore with a 10 nm $SiN_x$ amorphous film as the support, was used to form the vertical junctions with different twist angles. The ultra-thin $APbX_3$ perovskites on PDMS were used to stack on $SiN_x$ grid layer by layer through point-to-point transferring technique with the equipment shown in Supplementary Fig. 32.

**TEM imaging and spectrum acquisition.** The electron diffraction patterns and the HRTEM images were obtained on a 200 kV JEOL JEM-F200 field-emission microscope with a CMOS camara (Gatan Rio16) and an aberration-corrected JEOL GrandARM microscope equipped with a CMOS camera (Gatan OneView) and a direct electron detection camera (Gatan K2). A minimum dose system (MDS) was used to decrease the electron dose during searching and focusing the sample in order to achieve low-dose-rate imaging. Time series images of the same region of interest (ROI) were recorded and drift-corrected, and then superposed to obtain a HRTEM image with improved signal-to-noise ratio. In addition, energy dispersive X-ray spectroscopy (EDS) mapping (Supplementary Fig. 26-27) was done to show the elements distribution after collecting HRTEM images.

**Aberration-corrected STEM acquisition.** Scanning transmission electron microscopy (STEM) images were also obtained on the aberration-corrected JEOL GrandARM microscope under 300 kV with a convergence angle of 12 mrad at a dose around 114.7 $e/Å^2$.

**Analysis of periodic length in moiré superlattices.** The period lengths of these moiré superlattices could be calculated by $L = a_0/\sin\theta$, where $a_0$ is the length of in-plane Bravais lattice (6.3 Å for $MAPbI_3$ and 5.9 Å for $MAPbBr_3$) and $\theta$ is the twist angle. The calculated lengths are 13.6 nm for 2.5°, 5.0 nm for 7.3° and 2.3 nm for 15.0°, in excellent agreement with the results measured directly from the HRTEM images (**Fig. 2**).

**Computational Methods**

**DFT and TDDFT calculations.** All DFT calculations were performed using the Vienna Ab initio Simulation Package (VASP)[64] with projector augmented wave (PAW) potentials[65]. The generalized gradient approximation of Perdew, Burke and Ernzerhof (PBE)[66] was used for the exchange-correlation energy. The energy cutoff for the plane-wave basis was set to 400 eV. By cleaving along its (001) surface of the cubic $MAPbI_3$ crystal, we obtained a 2.5-layer-slab model (with a chemical formula of $MA_3Pb_2I_7$) to represent the halved $MAPbI_3$ crystal. The non-twisted $MAPbI_3$ bilayer was then constructed by overlaying the two identical slabs (AA stacking). The Brillouin zone was sampled using a $\Gamma$-centered $8 \times 8 \times 1$ Monkhorst-Pack grid for the non-twisted perovskite bilayers. The CellMatch software[67] was used to generate the atomic structures of two twisted $MAPbI_3$ bilayers with the twist angle $\theta = 16.5°$ and $\theta = 9.5°$. The moiré supercell lattice constants for the two twisted bilayers are: $a = b = 31.5$ Å ($\theta = 16.5°$) and $a = b = 37.8$ Å ($\theta = 9.5°$), respectively. Thanks to the large supercells (1650 and 2409 atoms), only the $\Gamma$ point was needed to determine optimized structures of the twisted bilayers. A vacuum length of ~12 Å was included in the out-of-plane direction to prevent spurious interaction between the periodic images. The force convergence threshold was set to 0.02 eV/Å. DFT-D2 correction[68] was used to account for vdW interaction between the organic cations and inorganic Pb-I frameworks. The 2D band structure was calculated along high symmetry k-point paths, including G or Γ at (0, 0, 0), X at (0.5, 0, 0), M at (0.5, 0.5, 0) and Y at (0, 0.5, 0).

The charge density of the lowest-energy exciton[69-70] was calculated using the linear-response time-dependent density functional theory (LR-TDDFT)[71] with optimally tuned, screened and range-separated hybrid exchange correlation functionals (OT-SRSH)[72-74]. There are three parameters α, β, and γ in the OT-SRSH functional with γ representing the range-separation parameter. And $\alpha$ and $\beta$ satisfy the equation: α + β = 1/ε₀, where ε₀ is the scalar dielectric constant. Herein, we used α = 0.05 and γ = 0.15 Å$^{-1}$ with ε₀ = 7.6 for the MAPbI₃ bilayers. Owing to the large moiré supercells, only the Γ-point was sampled in the Brillouin zone in the TDDFT-OT-SRSH calculations. Using these parameters, we estimate the exciton binding energies of the twisted MAPbI₃ bilayers as 0.16 and 0.13 eV for θ =16.5° and 9.5°, respectively. It is important to note that spin-orbit coupling (SOC) was not included in the DFT and TDDFT calculations owing to computational constraint. SOC is known to modify the conduction bands of MAPbI₃ which could lead to errors in the effective masses of electrons used in the microscopic theory for excitons (see below). However, the (flat) valence bands are not expected to be affected by the SOC.

**Calculations on excitonic properties.** We combine our DFT calculations with a fully microscopic many-particle theory in order to obtain direct access to the excitonic binding energies and wavefunctions, as a function of twist angle. Excitons in the perovskite heterojunctions can be described using the Wannier equation[75-76]:

$$\left(\frac{\hbar^2 \boldsymbol{q}^2}{2\mu_r} + E_G\right)\varphi_q^\mu + \sum_k V_{k+q}\varphi_k^\mu = E_\mu^b \varphi_q^\mu,$$

Here, $E_G$ is the band gap and $\mu_r$ is the reduced mass, both of which are obtained from the ab-initio DFT calculation. The Coulomb interaction, $V_q$, is modelled with the 2D-Keldysh interaction[76-77], taking into account the thickness of the layers (4 nm) as well as the screening of the layer and the substrate. We obtain a typical hydrogen-like series of exciton energy, $E_\mu^b$, and wave functions, $\varphi_k^\mu$. The lowest energy 1s exciton has a binding energy of around ~80 meV below the band-gap, in excellent agreement with our experiment. While DFT tends to underestimate the electronic band gap, the excitonic wavefunctions and binding energies are independent of the band gap.

Using our microscopic model, we can derive the so-called Elliot formula for the excitonic absorption[54]:

$$I_{PL} \propto \Im\left(\sum_\mu \frac{|M^\mu|^2 N_\mu^{Q=0}}{\hbar\omega - E_\mu^b + i\gamma^\mu}\right).$$

The PL intensity, $I_{PL}$, is therefore a series of Lorentzian peaks, centred on the exciton resonances, $E_\mu^b$, with broadening $\gamma^\mu$. This broadening is determined by the radiate and non-radiative decay channels in the perovskite, and can be extracted from the experimental line widths, which stays relatively constant with twist-angle. The angle-dependence of the PL intensity is therefore determined by the change in the excitonic optical matrix element, $M^\mu$, and the occupation, $N_\mu^{Q=0}$. At room temperature, the occupation effects are relatively small due to the large thermal distribution of excitons. Therefore, the change in the PL is driven mostly by the changing optical matrix element. The latter is directly proportional to the excitonic wave function:

$$|M^\mu|^2 \propto |\sum_k \varphi_k^\mu|^2 = |\psi^\mu(r=0)|^2.$$

where $|\psi^\mu(r=0)|^2$ can be understood as the probability of finding the electron and hole in the same place. As the bands flatten due to the changing twist angle, the excitons become more distributed in momentum space and hence more localised in real space leading to an increase in the optical matrix element and brighter PL emission.

The twist-angle dependent exciton diffusion can also be calculated microscopically. We use a many-particle approach, based on Wigner functions (see Ref. 77-79 for more details) to derive the temporal and spatial evolution of the exciton distribution. We arrive at an expression for the diffusion constant[76, 78]:

$$D = \frac{1}{2} \sum_{Q\mu} \frac{(v_Q)^2}{\Gamma_Q^\mu} \frac{exp^{-E_\mu^b(Q)/k_B T}}{Z}.$$

Here, $Z$ is the partition function, $v_Q$ is the exciton group velocity as a function of the centre-of-mass momentum, $Q$, and $\Gamma_Q^\mu$ is the exciton-scattering rate. While a full analysis of the internal scattering mechanisms which determine the exciton-scattering rate are beyond the scope of this work, we can estimate it using the assumption that $\Gamma_Q^\mu \approx \Gamma^\mu$, and is therefore related to the excitonic linewidth in the PL measurement[54, 76, 80], $\hbar\Gamma^\mu = \gamma^\mu$, which can be extracted from experiment.

Using these assumptions, we can plot the diffusion constant as a function of twist-angle (Fig. 5h). We obtain excellent qualitative agreement with the experiment and also very good quantitative agreement, especially at angles away from the 8°-10° range. Our model is likely to underestimate the diffusion at low twist angles, as the transport is less diffusive, moving towards a hopping regime[81].

**Device fabrication and measurement**

The Ti (4 nm)/Au (40 nm) electrode with the channel width of 10 μm was directly fabricated on SiO$_2$ (300 nm)/Si substrate using thermal evaporating method. Then, current–voltage (*I-V*) characteristics were acquired by sweeping the bias voltages from 1 V to -1 V with a step of 0.05 V based on a probe station (PS100 Lakeshore) with a source meter (Keithley 2400). The electrical conductivity was calculated as:

$$S = \frac{IL}{UwT}$$

Where $I$ is the current, $U$ is the voltage, $L$ is the channel length, $w$ is the channel width and $T$ is the thickness of samples. For $L$ and $w$, they are both 10 μm, and the thickness is defined by the AFM measurement.

**Data availability**

All data related to this study are available from the corresponding author on reasonable request.

**Acknowledgements** This work is primarily supported by the US Department of Energy, Office of Basic Energy Sciences under award number DE-SC0022082. The views expressed herein do not necessarily represent the views of the U.S. Department of Energy or the United States Government. L.D. acknowledges supports from National Science Foundation under award number 2143568-DMR. TEM work is supported by the Center for High-resolution Electron Microscopy (ChEM) at ShanghaiTech University and the Shanghai Key Laboratory of High-resolution Electron Microscopy. E.M. acknowledges support from the German Research Foundation (DFG) via the CRC 1083 (project B9) as well as via the regular project 504846924. G.L. acknowledges support from the US NSF PREM program (DMR-1828019) and the US Army Research Office (W911NF-23-10205).

**Author contributions** L.D. L.H. and S.Z. design the experiments. S.Z. synthesized and characterized all the 2D perovskite samples; L.J., D.B., S.K., Q.Z., and D.S. performed optical properties characterizations and data analysis; Y.L., B.Y. and Y.Y. performed TEM characterization and data analysis; J.P. and S.Z. did the confocal PL imaging on all the perovskite samples; L.Z., G.L., J.Y., J.T., E.M. and A.M.-K. performed DFT and microscopic simulations as well as data analysis; Y.L. and S.Z. performed device fabrication and characterization. B.F., A., Z.W. K.M. and H.P. participated in materials characterization and data analysis; S.Z., L.D. and L.H. wrote the manuscript; all authors read and revised the manuscript.

**Competing interests** The authors declare no competing interests.

**Additional information**

**Supplementary information** is available for this paper at XXX.

**Correspondence and requests for materials** should be addressed to L.H. or L.D.

**Reprints and permissions information** is available at http://www.nature.com/reprints.

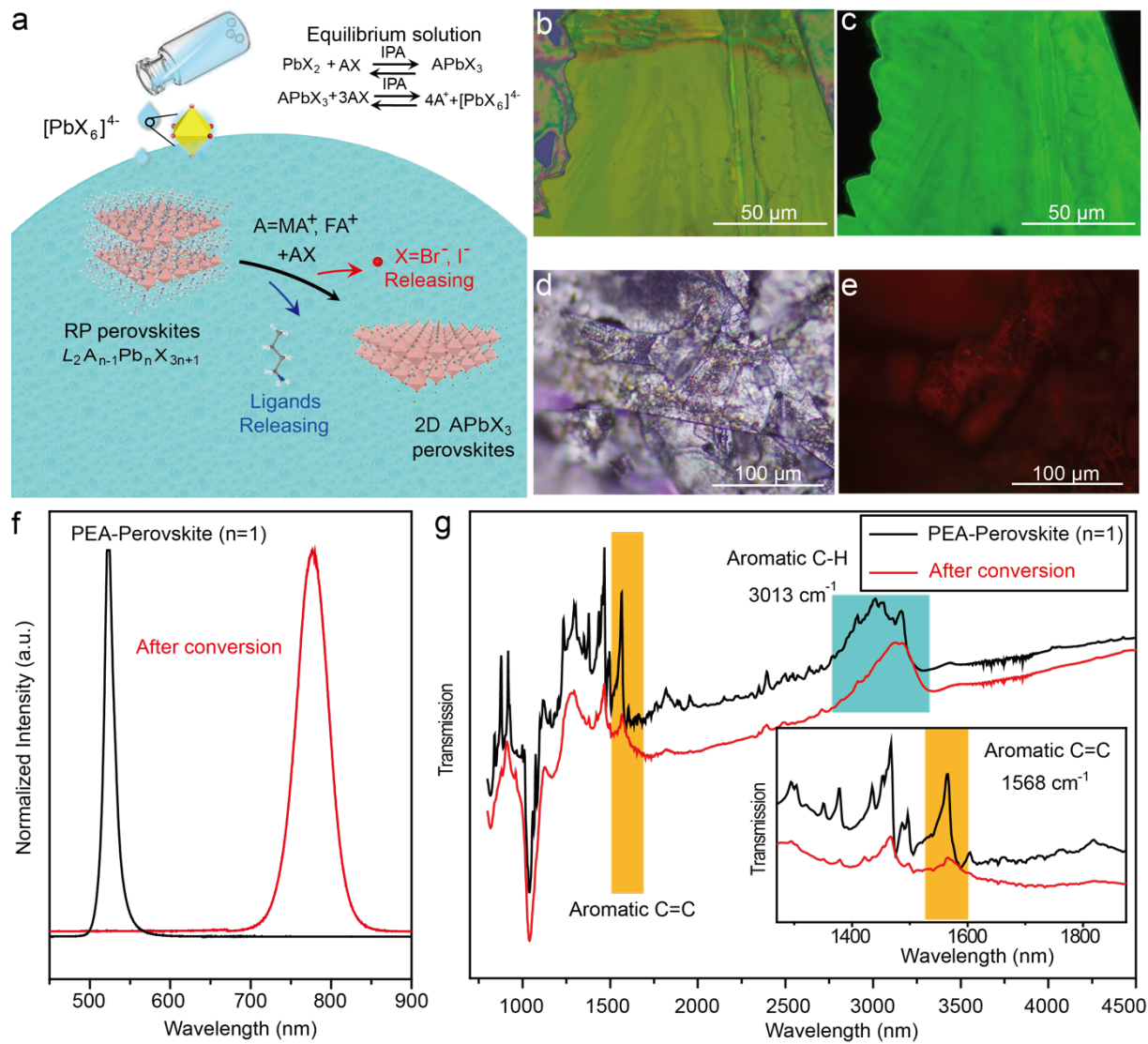

**Extended Data Fig. 1| IR characterization to prove the successful conversion to APbX₃ structure.**
**a**, Schematic to show the conversion process in IPA equilibrium solution using 2D RP phase perovskites as precursors. **b, c**, Optical (**b**) and PL (**c**) images of $(PEA)_2PbI_4$ single crystal. **d, e,** Optical (**d**) and PL (**e**) images of the crystal after conversion from $(PEA)_2PbI_4$ to $MAPbI_3$. **f,** PL emission spectrum of the crystals before and after conversion. **g,** IR spectrum of the crystals before and after conversion, and the disappearance of aromatic group indicates the loss of PEA ligand in crystals after conversion and successful conversion to $MAPbI_3$. Inset is the zoomed-in IR spectrum from 1280 nm to 1900 nm.

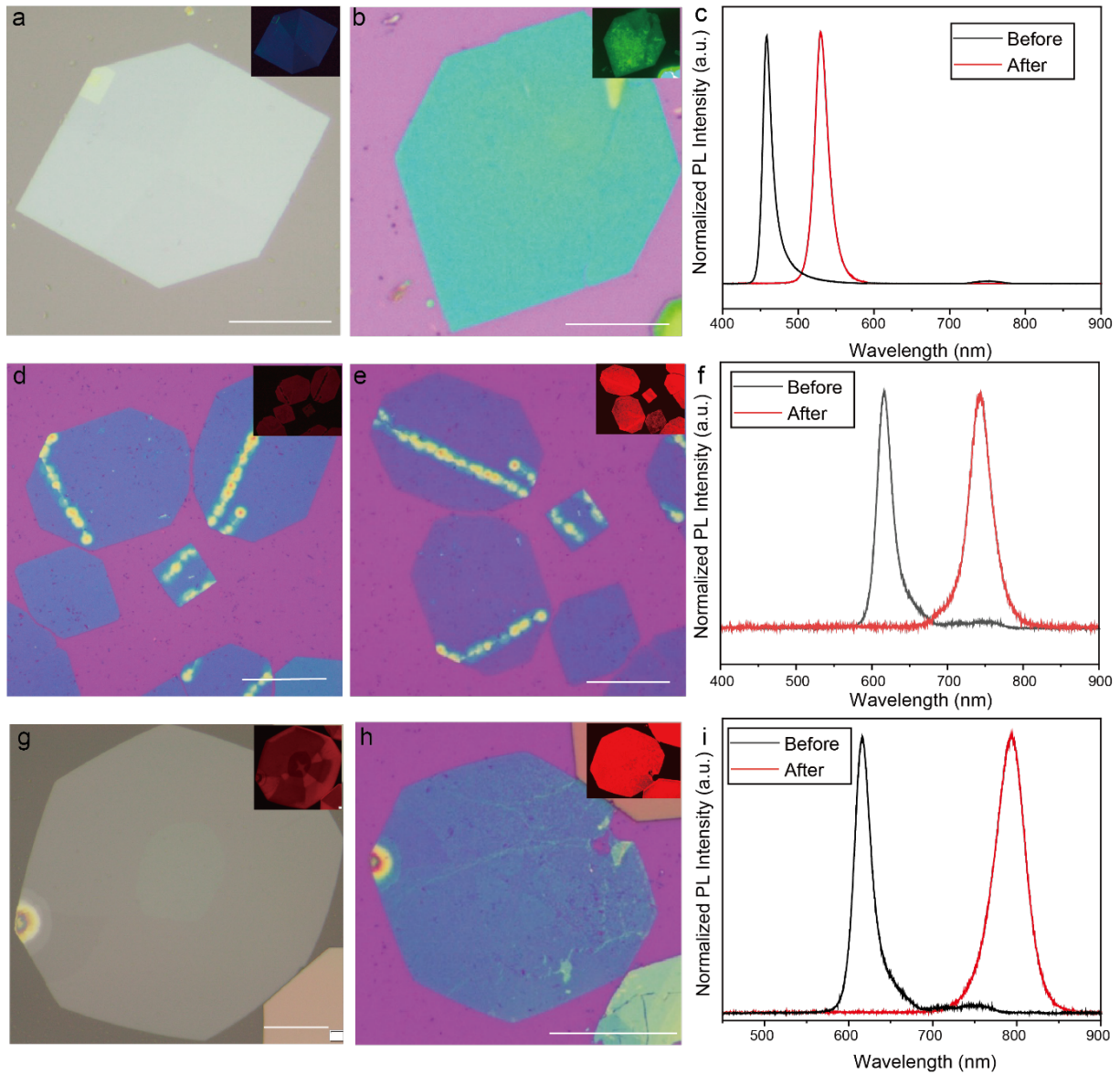

**Extended Data Fig. 2| Optical images and PL emission changes before and after conversion to APbX₃ perovskite. a, b,** Optical images of BA₂MA₂Pb₃Br₁₀ sheet before conversion (**a**) and after conversion to MAPbBr₃ (**b**), and their corresponding PL emission spectrum changes (**c**). The insets are the corresponding PL images. The scale bars are 20 μm in **a** and 10 μm in **b**. **d, e,** Optical images of BA₂MA₂Pb₃I₁₀ nanocrystal before conversion (**d**) and after conversion to MAPbI₃ (**e**), and the corresponding PL emission spectrum changes (**f**). The insets are the PL image for BA₂MA₂Pb₃I₁₀ sheet in (**d**) and confocal image for MAPbI₃ (the collecting wavelength from 730-760 nm) in (**e**). The scale bars are 50 μm in both **d** and **e**. **g, h,** Optical images of BA₂MA₂Pb₃I₁₀ sheet before conversion (**g**) and after conversion to FAPbI₃ (**h**), and the corresponding PL emission spectrum changes (**i**). The insets are the PL image for BA₂MA₂Pb₃I₁₀ sheet in **g** and confocal image for FAPbI₃ (the collecting wavelength from 780-800 nm) in **h**. The scale bars are 20 μm in **g** and 50 μm in **h**.

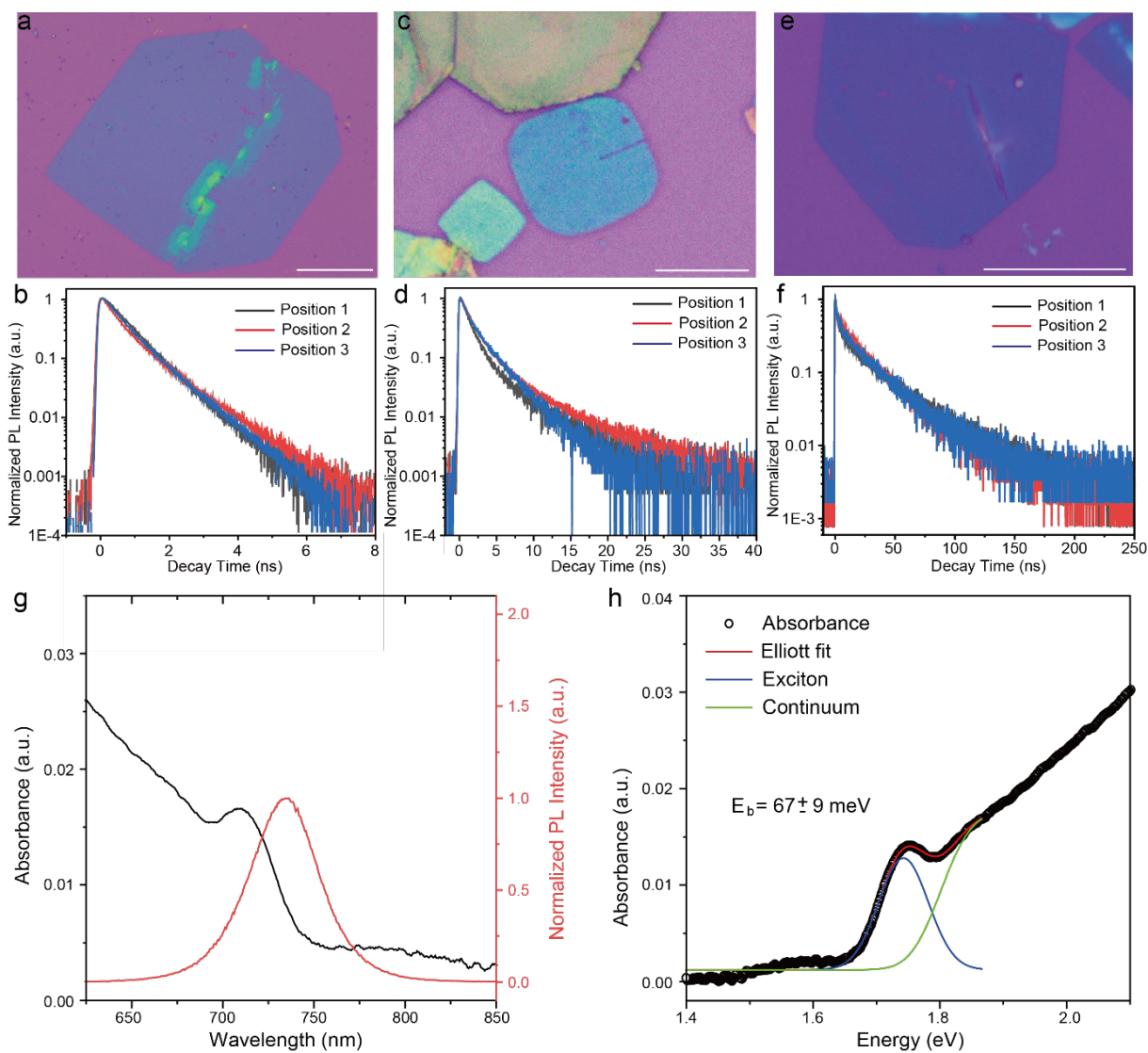

**Extended Data Fig. 3| Optical characterizations for different ultra-thin APbX₃ perovskite sheets.** **a**, **c**, and **e**, Optical images of 2D APbX₃ sheets, MAPbBr₃ (**a**), MAPbI₃ (**c**) and FAPbI₃ (**e**). **b**, **d**, and **f**, Corresponding lifetime characterizations in three positions, MAPbBr₃ (**b**), MAPbI₃ (**d**) and FAPbI₃ (**f**). The scale bars are 20 μm in **a**, 10 μm in **c**, and 20 μm in **e**. **g**, Absorption spectra (black line) and corresponding PL emission spectra (red line) of an ultrathin MAPbI₃ perovskite sheet. **h**, Corresponding fitting result (red line) based on Elliott model for the ultrathin MAPbI₃ perovskite sheet in **g**. The blue and green lines represent the excitonic and continuum contributions to the total absorption, respectively.

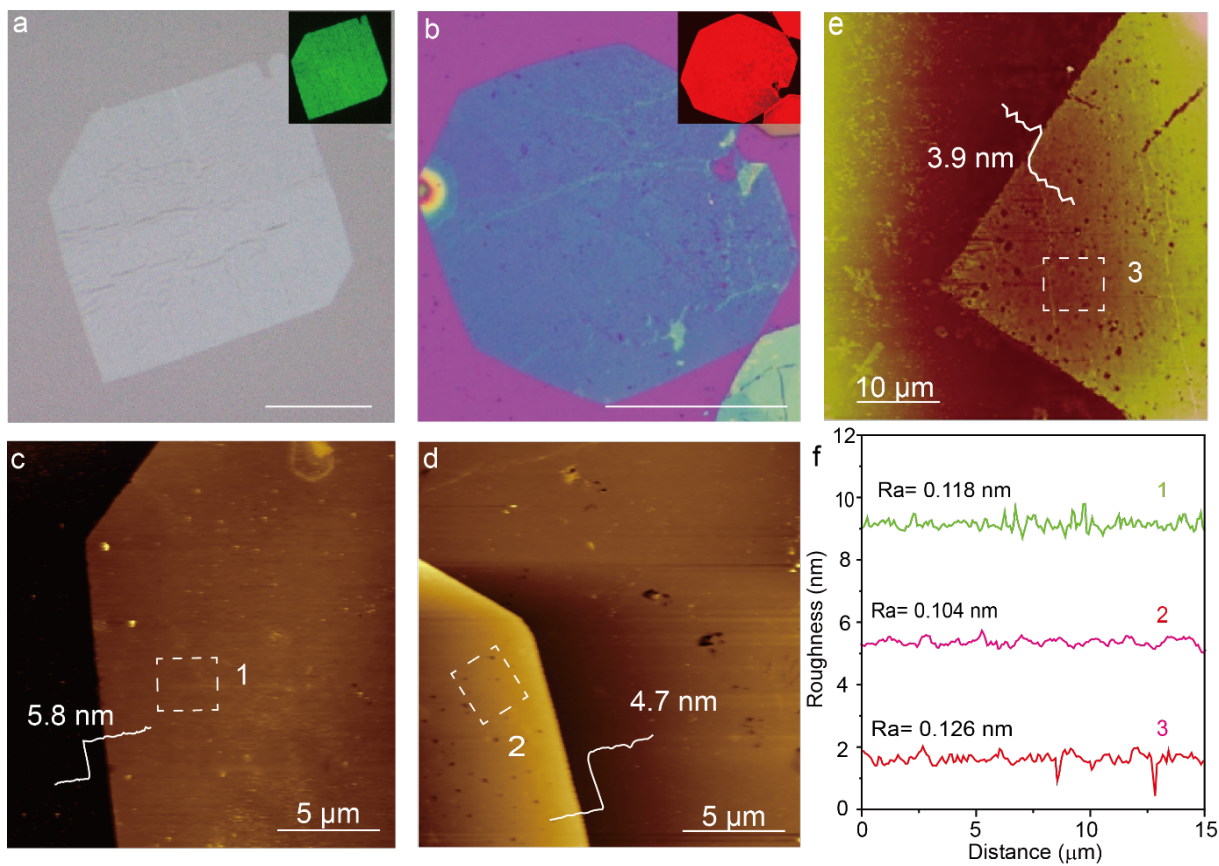

**Extended Data Fig. 4| More kinds of 2D APbX₃ perovskites and morphology characterization. a-b,** Optical images for MAPbBr₃ (**a**) and FAPbI₃ (**b**) ultra-thin ligand-free 2D perovskites. The inset images are the corresponding PL image for MAPbBr₃ and confocal image for FAPbI₃ (770-790 nm). The scale bars are 20 μm in **a** and 50 μm in **b**, respectively. **c, d, e,** AFM images of the ultra-thin APbX₃ perovskites, showing a thickness of ~5.8 nm for MAPbBr₃ (**c**), ~4.7 nm for FAPbI₃ (**d**) and ~3.9 nm for MAPbI₃ (**e**). **f,** Roughness analysis of APbX₃ ultrathin perovskite sheets corresponding to AFM images in (**c-e**).

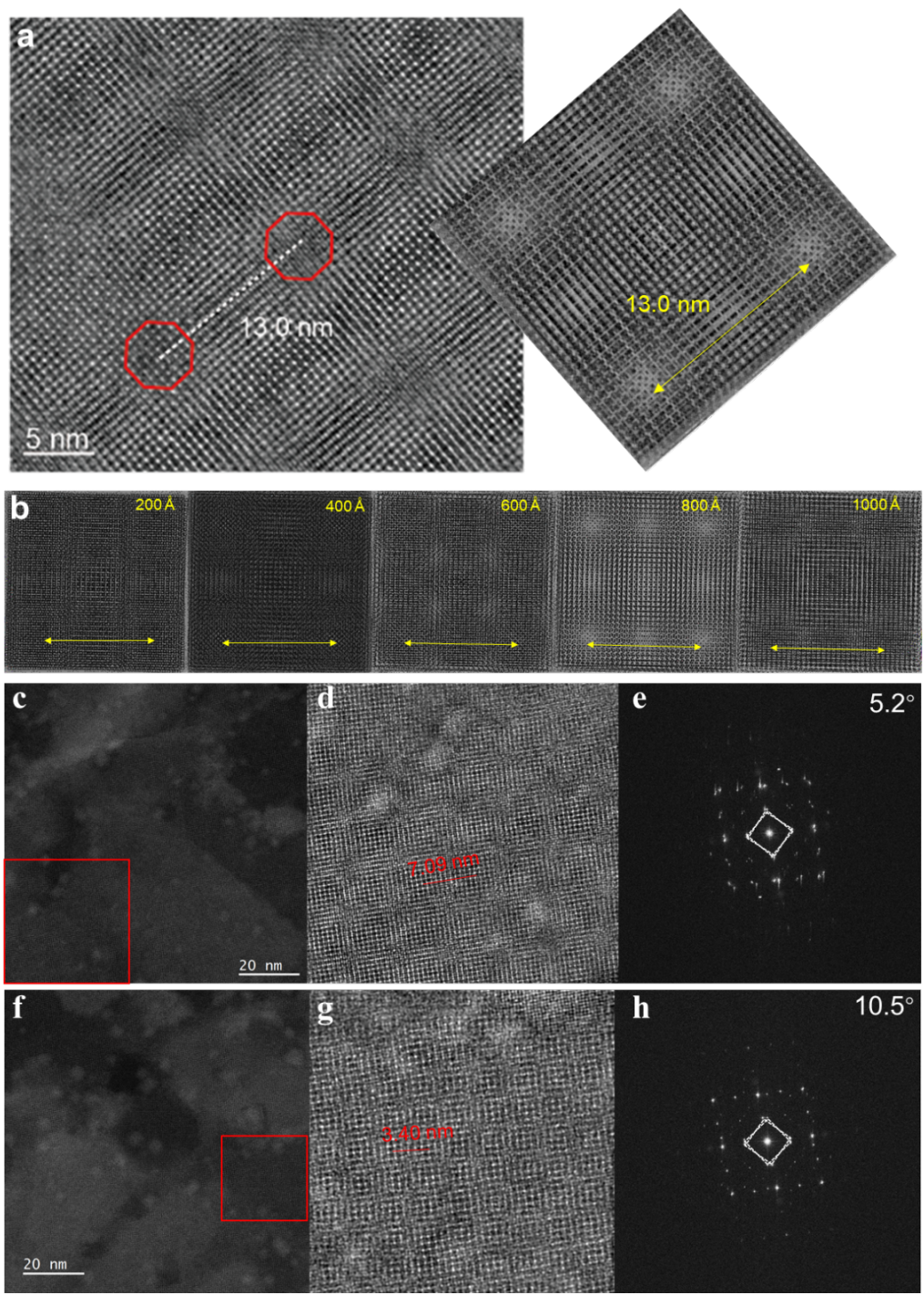

**Extended Data Fig. 5| Moiré fringes in TEM and STEM modes. a**, The TEM image from **Fig. 2d** in the manuscript, overlaid with a simulated TEM image, showing the same moiré period of 13.0 nm. **b**, Simulated TEM images under different defocus ranging from 200 to 1000 angstrom, showing the same moiré period. **c, f,** STEM images of two MAPbI$_3$ TPLs with different twisting angles. **d, g,** Enlarged Wiener filtered images in the red box in **c** and **f**, indicating the periodic length of 7.09 nm and 3.40 nm. **e, h,** Corresponding FT patterns of moiré patterns in the two MAPbI$_3$ TPLs, showing the twisting angles are 5.2° for **c** and 10.5° for (**f**), respectively.

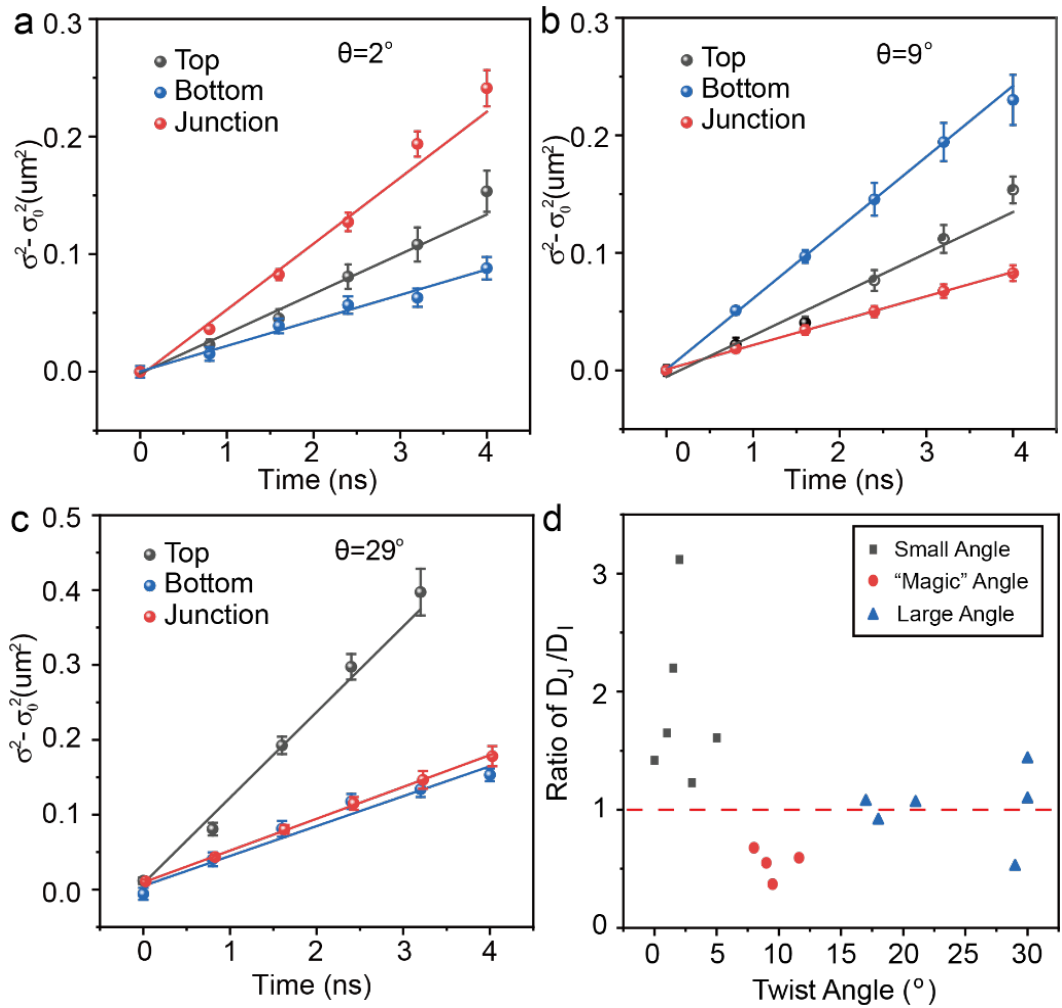

**Extended Data Fig. 6| Detailed angle-dependent exciton diffusion in MAPbI₃ TPLs. a, b, c,** Determining the corresponding exciton diffusion coefficient by fitting t-$\sigma_t^2$ at junction parts in MAPbI₃ TPLs with different twist angles of ~2° (**a**), ~9° (**b**) and ~29° (**c**). **d,** Detailed statistics on exciton diffusion of TPLs. $D_J$ means the carrier diffusion at junction part and $D_I$ indicates the mean value of individual top and bottom perovskites in each stacking structure.

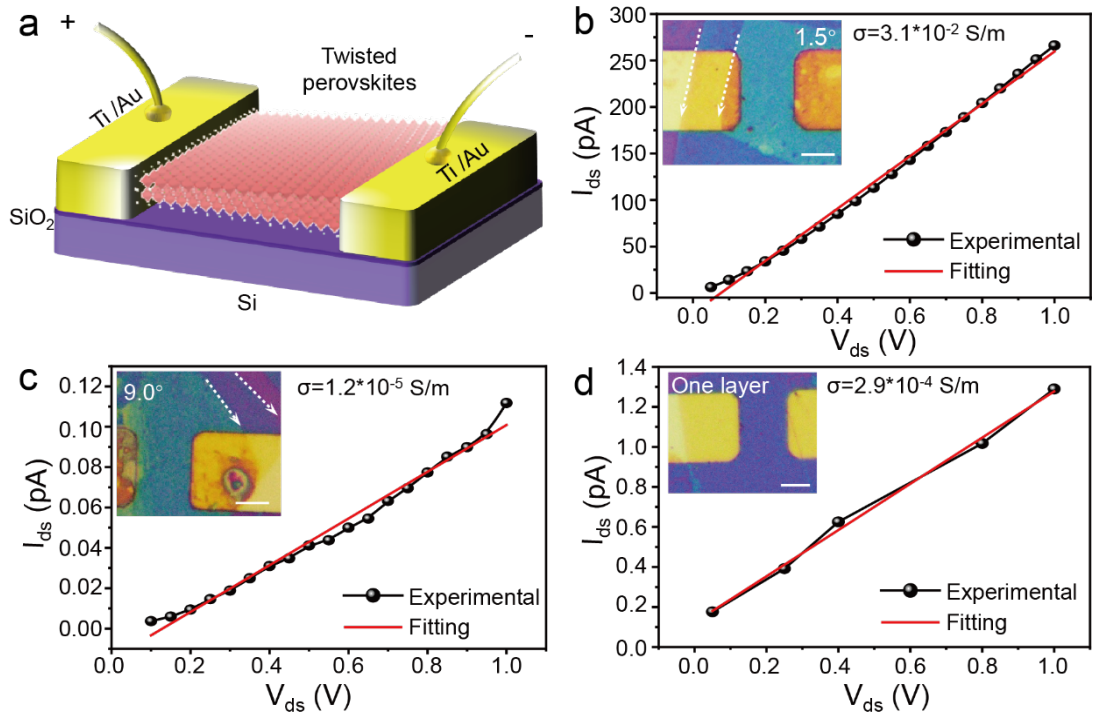

**Extended Data Fig. 7| Device characterization for MAPbI₃ TPLs. a,** Device structure of MAPbI₃ TPLs with Au as electrodes. **b,** *I-V* characteristics in dark condition for MAPbI₃ TPLs of 1.5°, and the inset is the optical image of corresponding device with the channel width of 10 μm. The scale bar is 5 μm. **c,** *I-V* characteristics in dark condition for MAPbI₃ TPLs of 9.0°, and the inset is the optical image of corresponding device with the channel width of 10 μm. The scale bar is 5 μm. **d,** *I-V* characteristics in dark condition for individual MAPbI₃ perovskite layer, and the inset is the optical image of corresponding device with the channel width of 10 μm. The scale bar is 5 μm.

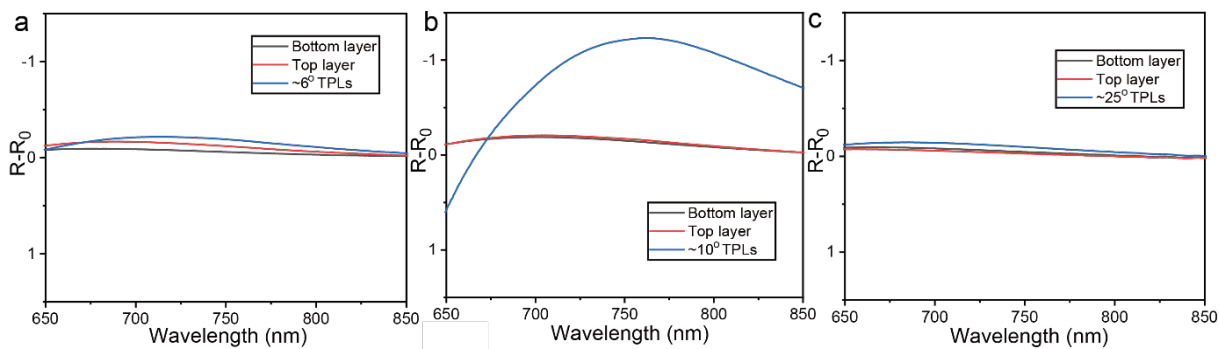

**Extended Data Fig. 8| Reflectance difference spectra and calculated density of states (DOS) of TPLs**. **a, b, c**, Reflectance difference spectra measured near the exciton energy in individual MAPbI₃ regions (black, red) and in twisting region (blue) with various twist angles, ~6° (**a**), ~10° (**b**) and ~25° (**c**). For details of the reflectance difference experiments, in particular parameters R and R₀, see Methods.

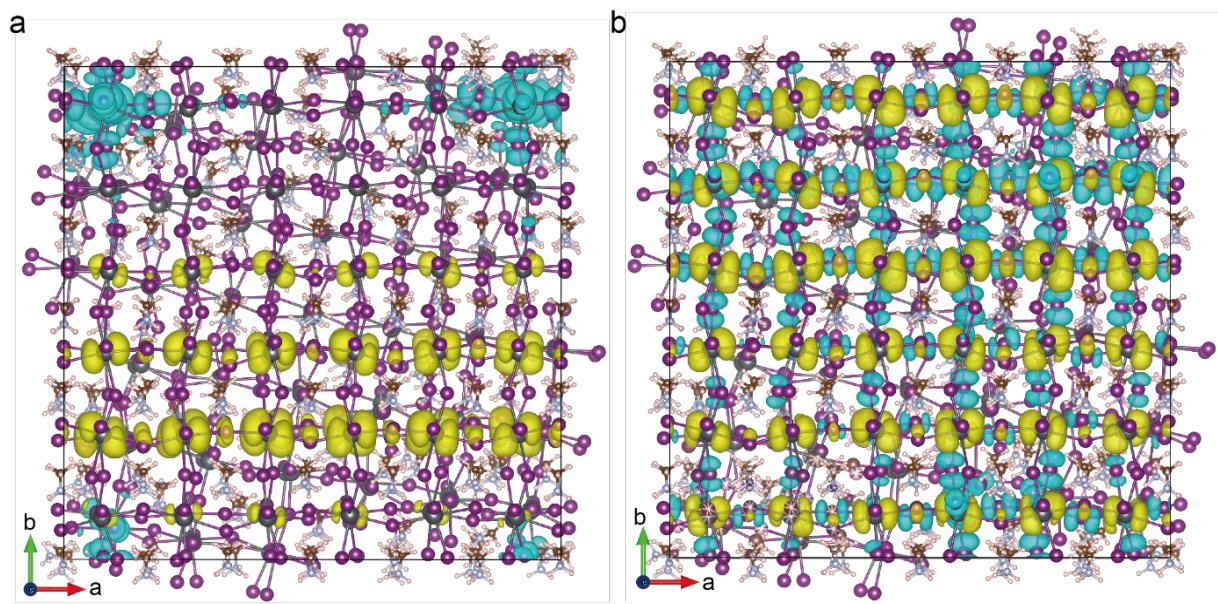

**Extended Data Fig. 9| The charge density of the lowest energy exciton in TPLs. a,** The charge density in TPLs with the twist angle of 9.5°. **b,** The charge density in TPLs with the twist angle of 16.5°. The hole and electron densities (iso-surface value set at 0.0005 eÅ$^{-3}$) are shown in cyan and yellow color, respectively.

**Extended Table 1. Current methods to realize APbX$_3$ perovskites and comparison on potential to form twisted structure**

| Perovskite species | Synthesis methods | Domain size | Thickness | Reaction Temperature | Comments on fabricating twisted structures |
|---|---|---|---|---|---|
| MAPbI$_3$[38] | CVD | 9 μm | 1-13 nm | 120 °C | Only on solid substrates; cannot be transferred* |
| MAPbI$_3$[S11] | CVD conversion | 10 μm | 50-240 nm | 380°C (PbI$_2$) 120 °C (conversion) | Only on solid substrates; too thick; cannot be transferred |
| FAPbI$_3$[S12] | CVD conversion | 1-5 μm | 4 nm | 360°C (PbI$_2$) 150 °C (conversion) | Only on WS$_2$; small domain; cannot be transferred |
| MAPbCl$_3$[S13] | CVD | 40 μm | 9 nm | 220°C | Only on mica substrate; a little thick; cannot separate it from mica completely |
| FAPbI$_3$[41] | Gas conversion | 10 μm | 8-20 nm | 180°C (PbI$_2$) 90-95 °C (conversion) | Only on silicon substrate; a little thick; cannot be transferred |
| MAPbI$_3$[41] | Gas conversion | 10 μm | 4-20 nm | 180°C (PbI$_2$) 130-150 °C (conversion) | Only on silicon substrate; a little thick; cannot be transferred |
| CsPbBr$_3$[S14] | Solution synthesis | 5 μm | 3 nm | 50-150 °C | Small domain; hard to get clean samples and further fabricate |
| APbX$_3$ in our paper | Our equilibrium solution method | 5~200 μm | 2~100 nm | Room temperature | On any substrate; controlled domain size; controlled thickness |

*Note: For stacking two thin sheets together, it is necessary to firstly transfer one layer onto a transparent and flexible substrate (such as PDMS or PMMA). However, this process typically involves etching and lift-off with polar solvents. This process is widely used in other 2D materials but unfortunately it can damage the fragile perovskite sheets or lead to incomplete transfer. Therefore, it is necessary to directly grow one 2D perovskite sheet directly on a soft polymer substrate, but this has not been realized before. As shown above, none of previously reported methods is suitable for fabricating twist angle-controlled 2D perovskite homojunctions.

**Extended Table 2. Exciton recombination rate constants and exciton annihilation rate constants for MAPbI$_3$ TPLs with different twist angles.**

| | TPLs (~2°) | | | TPLs (~9°) | | | TPLs (~29°) | | |
|---|---|---|---|---|---|---|---|---|---|
| | Top | Bottom | Junction | Top | Bottom | Junction | Top | Bottom | Junction |
| $k_1$ (ns$^{-1}$) | 0.08± 0.01 | 0.10 ±0.01 | 0.07 ±0.01 | 0.18 ±0.01 | 0.25 ±0.01 | 0.18 ±0.02 | 0.22 ±0.01 | 0.08 ±0.02 | 0.15 ±0.02 |
| $k_2$ (10$^{-3}$ cm$^2$s$^{-1}$) | 0.74 ±0.07 | 1.2 ±0.03 | 2.2 ±0.1 | 2.5±0.2 | 3.5 ±0.2 | 0.045 ±0.05 | 1.6 ±0.1 | 6.0 ±0.2 | 2.1±0.1 |

Note: $k_1$ and $k_2$ are the exciton recombination rate and exciton annihilation (EEA) rate, respectively. The data are extracted from the fitting shown in **Fig. 3** and Supplementary Fig. 43. The variation on $k_1$ and $k_2$ values of individual layers is due to thickness difference. It's worth noting that $k_2$ for smaller and larger twist angles generally falls within the range of the individual layers. In contrast, $k_2$ is reduced by two orders of magnitudes for twisted angle of 9°, which can be explained by the localized excitons from the flat bands.